\RequirePackage{fix-cm}
\documentclass[smallextended]{svjour3}       
\smartqed  
\usepackage{graphicx,amssymb}
\usepackage{dcolumn}
\usepackage{bm}
\usepackage{amssymb}
\usepackage{amsmath}
\usepackage{amsfonts,mathrsfs}

\begin{document}

\title{ Lorentz-invariant, retrocausal, and deterministic hidden variables}


\author{Aur\'elien Drezet}


\institute{A. Drezet \at Institut NEEL, CNRS and Universit\'e Grenoble Alpes, F-38000 Grenoble, France \\
            \email{adrezet@neel.cnrs.fr} }

\date{Received: date / Accepted: date}

\maketitle

\begin{abstract}
We review several no-go theorems attributed to Gisin and Hardy, Conway and Kochen purporting the impossibility  of  Lorentz-invariant deterministic hidden-variable model for explaining quantum nonlocality. Those theorems claim that the only known solution to escape the conclusions is either to accept a preferred reference frame or to abandon the hidden-variable program altogether. Here we present a different alternative based on a foliation dependent framework adapted to deterministic hidden variables.  We analyse the impact of such an approach on Bohmian  mechanics and show  that retrocausation  (that is future influencing the past) necessarily comes out without time-loop paradox.      

\end{abstract}

\section{Introduction}\label{sec1}
\indent Quantum nonlocality (QN), as demonstrated by Bell's theorem~\cite{Bell}, is certainly a cornerstone scientific discovery of the last century. As such it motivated and renewed the full field of research about quantum foundations and pushed researchers forward to develop quantum protocols and algorithms with huge potential technological applications. However, appreciation of QN physical implications and importance strongly fluctuates from specialist to specialist.\\
\indent One of the central issue in this debate concerns the constraints imposed on the underlying hidden variables or beables which could possibly explain QN through a mechanical description. Probably the most popular hidden-variable model is the de Broglie-Bohm pilot-wave interpretation~\cite{deBroglie,Bohm} popularized under the name Bohmian mechanics (BM) and which is notoriously nonlocal (i.e., involving faster-than light influences) and contextual. BM is a deterministic approach which fully reproduces quantum mechanical probabilistic predictions (i.e., BM is empirically equivalent to the standard view). BM  also motivated  Bell proof's that QN, that is the negation of causal Einstein locality, is unavoidable in any discussion concerning hidden variables \cite{Bell2}. However, despite this success in the non relativistic regime there is no yet consensus on the possible extension and generalization of BM to the relativistic level, and the theory suffers from the drawback of being manifestly non-Lorentz invariant or covariant. \\
\indent Bohm and also Valentini and Bell~\cite{Bohm,ValentiniPhd,Bell,Ghost} already favored the view admitting a preferred space-time foliation or reference frame acting as a kind of new Aether (sometimes called the `sub-quantum' Aether~\cite{Bohm,Vigier1,Vigier2}). For many this perspective would seriously represent a tentative of regression to the pre-Einsteinian era when Lorenz and others proposed a mechanical explanation of electromagnetic phenomena in a preferred reference frame. An Aether would indeed violate the mere spirit of special relativity principle which puts on an equal level the description in every Lorentz frames. Admittedly, it also justifies many doubts and criticisms concerning the physical plausibility of any BM relativistic extension along this line.\\
\indent Over the years several researchers have attempted a mathematical demonstration, similar in philosophy to Bell's theorem, which would prohibit the mere existence of covariant nonlocal hidden-variables. Among the various  works in this direction Hardy's nonlocality without inequality proof \cite{Hardy1a,discussion} was key by emphasizing  the role  of Lorentz-invariant `elements of reality', (i.e., independent of any reference frame) and counterfactual reasoning involving  different inertial observers in relative motions. This prompted several subsequent analysis by Hardy~\cite{Hardy2}, Conway and Kochen~\cite{Kochen1,Kochen2}, and more recently Gisin~\cite{Gisin1,Gisin2} and Blood~\cite{Blood} resulting into no-go theorems against the existence of covariant nonlocal deterministic or stochastic hidden variable approaches.  In particular, the work by Conway and Kochen~\cite{Kochen1,Kochen2} (i.e., the so called `Free-will theorem') stirred  important controversies~\cite{discussion2a,discussion2b,discussion2c} about theories involving nonlocal stochastic hidden variables (e.g., the spontaneous collapse `GRW-flash' ontology proposed by Tumulka~\cite{Tumulka,GRW}).\\ 
\indent In the present work I want to go back to the claims surrounding the purported non-existence of covariant nonlocal deterministic hidden variables \cite{Hardy2,Gisin1,Blood,Kochen1,Kochen2}. My purpose is to show that it exists at least one way to bypass these no-go theorems and thus to define a Lorentz invariant extension of deterministic hidden-variable theories a la de Broglie Bohm. Remarkably, our result is very robust  and can be implemented in various scenarios such as relativistic and covariant BM. Also, to paraphrase Gisin~\cite{Gisin1} while our analysis is based on a pretty simple reasoning (hence possibly well known to some readers) the present discussion will bring  to the attention of the community some essential properties of QN which are necessary in order to build up a satisfying Lorentz invariant deterministic quantum ontology. 
\section{Reviewing the no-go theorems}\label{sec2}
\indent We start with reviewing  Hardy's  condition for the Lorentz-invariance of hidden-variable theories (LIHVT): \begin{quote}\textit{For a given run of an experiment (i.e. for a given set of hidden variables), a hidden-variable theory must give the same predictions for outcomes of measurements, regardless of the frame of reference $\mathscr{F}$ in which it is applied}~\cite{Hardy2}. \end{quote}  In a classical but relativistic context such a condition is  natural since particle or field trajectories are the invariant, i.e., absolute objects  of the theory which are univocally defined  through  their space-time evolutions.  A  system of $N$ point-like particles is for example described by space-time coordinates $x_i^{\mu_i}(s_i)\in \mathbb{R}^4$  (with $i=1,...,N$ and $\mu_i=0,...,3$) defining  $N$ curves parametrized by $s_i$ (which can be the proper-times of each particles). Different observers in relative motions would see the trajectories differently but  all these relative views refer to the same objects related by Lorentz (or more general space-time) transformations, i.e., in agreement with Einstein's relativity.\\
\indent The difficulty to extend this kind of space-time ontology to the quantum regime is untimely linked to QN acting between particles through space-like intervals and therefore conflicting with our usual notions of causality and time-ordering for events.\\ 
\indent More precisely, the usual causality would intuitively impose the following principle of outcome independence from later measurements (POILM):   
\begin{quote}\textit{When the
hidden-variable theory is applied in a reference frame $\mathscr{F}$, then the outcome of a measurement Q made
at time t does not depend on the choice of what is
measured at times later than t (viewed in frame $\mathscr{F}$)
even if these later measurements are made in a region
separated from the region in which measurement Q
is made by a space-like interval. }~\cite{Hardy2}. \end{quote} 
\indent As explained by Hardy this principle naturally follows from the assumption that the hidden-variable description of two systems should be disjoints when the systems are uncorrelated (i.e., when the quantum state can be  written as a product)~\cite{Hardy2}. Indeed, consider a situation where quantum measurements are realized in two disjoint spacetime regions A and B on a system S.  Before the measurement  at time $t_0$ (in a reference frame $\mathscr{F}$) the quantum state is a product $|\textrm{Alice}_0\rangle|S_0\rangle|\textrm{Bob}_0\rangle$  where $|\textrm{Alice}_0\rangle$, $|\textrm{Bob}_0\rangle$ denote the measurement apparatuses quantum states, $|S_0\rangle$ the state of S, and we thus suppose that we can define three disjoint sets of hidden variables  $\lambda_{\textrm{Alice}}$, $\lambda_S$, and $\lambda_{\textrm{Bob}}$. Now at time $t_A>t_0$ the local measurement at A entangle the sub-systems Alice+S and we get the new state:
\begin{equation}
(\sum_i c_i|\textrm{Alice}_i\rangle|S_i\rangle)|\textrm{Bob}_0\rangle
\end{equation}  where  $|\textrm{Alice}_i\rangle$,$|S_i\rangle$  are states available to Alice+S. However, the system Bob is still factorized and  thus independent of Alice+S.   Clearly, if we now at time $t_B>t_A$ apply the  measurement at B  we get a new entangled quantum state for  Alice+S+Bob:
 \begin{equation}
\sum_{i,j} c_id_{j,i}|\textrm{Alice}_i\rangle|S_{i,j}\rangle|\textrm{Bob}_j\rangle.
\end{equation} From the perspective of the hidden-variable theory we thus expect that the outcomes $\alpha$ and $\beta$ of the two consecutive measurements at A and B are defined by functions 
\begin{eqnarray}
\alpha=f_\Psi(\textbf{a},\lambda_S, \lambda_{\textrm{Alice}}), & 
\beta=g_\Psi(\textbf{a},\textbf{b} , \lambda_S,\lambda_{\textrm{Alice}},\lambda_{\textrm{Bob}})
\end{eqnarray} where $\textbf{a}$ and, $\textbf{b}$  are some local settings for the two measurements. From the notations it is obvious that  the measurement and hidden variable  evolution at  A can not depend on what will later be done in region B.  Inversely, $\beta$ can nonlocally depend on all variables and this even if the  regions A and B are space-like separated.  This issue summarizes the contents of POILM. \\
\indent Moreover, POILM fits well with the standard quantum formalism in which the unitary evolution through time (i.e., associated  with the first order differential Schrodinger equation $i\frac{d}{dt}\Psi(t)= H\Psi(t)$) defines unambiguously the quantum state $\Psi(t+\Delta)$ at time $t+\Delta$ knowing the quantum state $\Psi(t)$ at an earlier time $t$. In this first-order dynamics, the Cauchy problem, i.e., the  evolution of the wave function in the future is univocally determined by the knowledge about the quantum state in the past. In turn, retrocausation (i.e. the future influencing the past) is avoided from the wave function evolution since a later measurement can not influence an earlier one~\footnote{We emphasize that this doesn't contradict time-symmetry of the unitary evolution: it is indeed possible to describe univocally the wavefunction in the past knowing the quantum state in the future.  }. POILM postulates that this must also be true  at the hidden-variable level as for example in non-relativistic Bohmian mechanics. This freedom of choice concerning the future of the system described by $\lambda_S$ entails therefore a kind of free-will~\cite{Kochen1,Kochen2} and an absence of super-determinism which would otherwise couple $\lambda_S$, $\lambda_{\textrm{Alice}}$, and $\lambda_{\textrm{Bob}}$ (i.e. we assume $\lambda$-independence).\\ 
\indent Actually, it is the simultaneous application of POILM and LIHVT to quantum entangled systems which leads to some fundamental contradictions and thus results into the above mentioned no-go theorems about covariant nonlocal deterministic hidden-variables~\cite{Hardy2,Gisin1,Blood}.\\ \indent Consider, for example the no-go theorem by Gisin and Blood~\cite{Gisin1,Blood} where the two atoms of an entangled pair prepared in a singlet EPR state $\Psi^{(-)}$ are spacelike separated
\begin{figure}[hbtp]
\begin{center}
\includegraphics[width=0.65\textwidth]{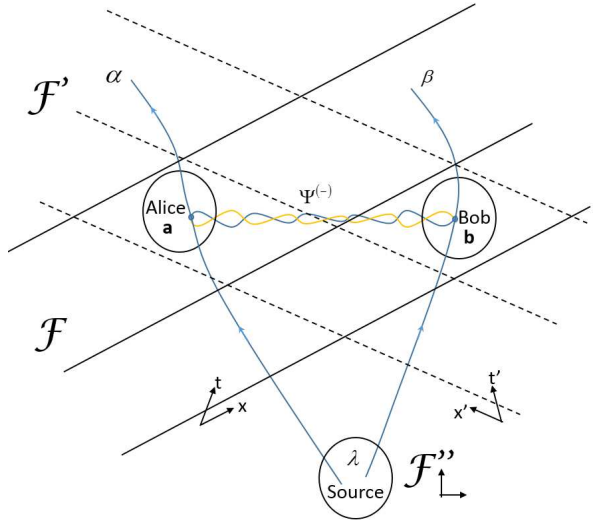}\\
\caption{Principle of Gisin's paradox.  }
\label{fig1}
\end{center}
\end{figure}
 in two regions where agents Alice and Bob record their spins using Stern and Gerlach measuring settings. Since Alice and Bob have the freedom to select the directions of the spin-analyzer settings (labeled $\textbf{a}$ and $\textbf{b}$) we have here the complete scenario leading to Bell's inequality. Now,  as illustrated on Fig.~\ref{fig1},  we can define a Lorentz reference frame $\mathscr{F}$ (associated with a given hyperplane foliation $\mathcal{F}$ of space time defining time leaves $t=const.$) such that the detection by Bob leading to the outcome $\beta$ occurs before the detection by Alice of the outcome $\alpha$. Following POILM we thus expect causal relations
\begin{eqnarray}
\beta=F_{\textrm{BA}}(\Psi^{(-)},\textbf{b},\lambda_{\textrm{Bob}},\lambda_S), &
\alpha=S_{\textrm{BA}}(\Psi^{(-)},\textbf{a},\textbf{b},\lambda_{\textrm{Alice}},\lambda_{\textrm{Bob}},\lambda_S)\label{1}
\end{eqnarray}  where $F_{\textrm{BA}}$ and $S_{\textrm{BA}}$ are two functions depending on the hidden variables $\lambda_{\textrm{Alice}}$, $\lambda_{\textrm{B}}$, and $\lambda_S$ (e.g., defined in the remote past). Causality implies that $\beta$ can not depends on $\textbf{a}$ (specifically if the  basis choice by Alice and Bob is decided at the last moment through some random mechanisms). Still, $\alpha$ can depend of both $\textbf{a}$ and $\textbf{b}$  settings through some nonlocal interactions. However, from the point of view of a second Lorentz observer $\mathscr{F'}$ (associated with a hyperplane foliation $\mathcal{F'}$ defining time leaves $t'=const.$) the time sequence is reversed (see Fig.~\ref{fig1}) and  from POILM we should expect instead the causal relation      
\begin{eqnarray}
\alpha=F'_{\textrm{AB}}(\Psi^{(-)},\textbf{a},\lambda_{\textrm{Alice}},\lambda_S),&
\beta=S'_{\textrm{AB}}(\Psi^{(-)},\textbf{a},\textbf{b},\lambda_{\textrm{Alice}},\lambda_{\textrm{Bob}},\lambda_S)\label{2}
\end{eqnarray} with obvious notations.
 The contradiction with quantum mechanics arises when one is applying POILM together with LIHVT since by equaling the outcomes of Eqs.~\ref{1} and \ref{2} we should get as a result that  $S_{\textrm{BA}}$ (respectively $S'_{\textrm{AB}}$) is independent of $\textbf{b}$ and $\lambda_{\textrm{Bob}},$  (respectively $\textbf{a}$ and $\lambda_{\textrm{Alice}}$). Therefore, in this scenario we end up with a local model  which necessarily  violates Bell's inequality in blatant contradiction with the pre-requirement.\\
\begin{figure}[hbtp]
\begin{center}
\includegraphics[width=0.8\columnwidth]{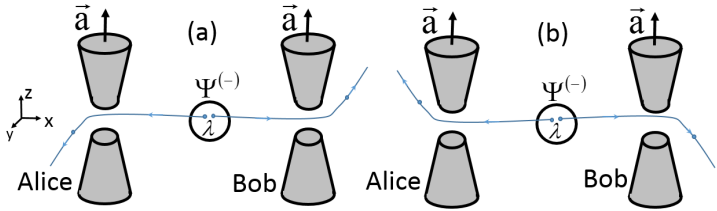}\\
\caption{ Gisin's paradox within a Bohmian perspective in two Lorentz frames (a) and (b).  }
\label{fig2}
\end{center}
\end{figure}
\indent As an illustration of the consequence of this no-go theorem  we show on Fig.~\ref{fig2} the Bohmian description given by Bricmont~\cite{Bricmont} of the EPR-Bell paradox in the case where Alice and Bob settings are the same (i.e., $\textbf{a}=\textbf{b}$). Supposing that in Earth common reference frame $\mathscr{F''}$ (assumed to be Lorentz) the source is exactly in between the two settings then Alice and Bob will detect simultaneously the particles. With the singlet state $\Psi^{(-)}$ perfect anti-correlation will naturally arise and Alice and Bob shall always observe opposed outcomes.  Now, suppose that we apply BM in the Lorentz reference frame $\mathscr{F}$ moving to the right in the $+x$ direction (i.e. with a constant positive velocity $v_x=v_e>0$). In other words we use $\mathscr{F}$  and the foliation $\mathcal{F}$ to compute the Bohmian paths.  In $\mathscr{F}$ Bob is detecting the spin before Alice.  Exploiting the symmetry of the initial wave-packet \footnote{In BM particle trajectories can not cross in the configuration space: this play a key role in the deduction~\cite{Bricmont,Rice}.} and QN~\cite{Bricmont,Rice} we can always find Bohmian trajectories such that for a given $\lambda$ (i.e., the  initial space-time coordinates of the pair) the particle detected by Bob leads to the outcome $\beta=+1$ while  Alice 
detects her spin outcome along $\alpha=-1$ (see Fig.~\ref{fig2}(a)). However, the result is reversed if we evaluate the Bohmian paths from the Lorentz frame $\mathscr{F}'$ moving to the left with the velocity $v'_x=-v_e<0$. In $\mathscr{F}'$  for the same initial conditions $\lambda$  Alice gets her result first  with the value $\alpha=+1$ and then nonlocality forces Bob outcome to the value $\beta=-1$ (see Fig.~\ref{fig2}(b)). Obviously, both nonlocal descriptions can not be true at once. Therefore, one have to make a choice and there is apparently a kind of preferred picture or foliation $\mathcal{F}$ or $\mathcal{F}'$ involved in BM to be non contradictory with special relativity.  This is one of the conclusion on which we are apparently forced upon if we take seriously the Gisin-Blood no-go theorem~\cite{Gisin1,Blood}. Indeed, if we want to preserve nonlocality at the hidden-variable level without contradicting the theorem we have to relax or abandon one of the hypothesis LIHVT and POILM. Relaxing or amending LIHVT would be very constraining on hidden variables since it would contradict the mere spirit of classical physics which is to obtain a causal description in space-time independently of the reference frame chosen. Also already in standard quantum mechanics joints probability are Lorentz invariant observable and it seems natural to suppose that it should so at the hidden-variable level. The strategy taken by BM is to restrict the application of POILM to a preferred reference frame $\mathscr{F}_0$ in which the trajectories can be evaluated. One can still apply LIHVT and transform the particle paths in a different Lorentz frame $\mathscr{F}$ but the privileged foliation $\mathcal{F}_0$ will acts as a kind of Aether to which we have to go back to compute the trajectories from the wave function.\\ 
\indent A different deduction is obtained by Conway and Kochen~\cite{Kochen1,Kochen2} with their particular version of the theorem based on perfect entanglement between two spin 1  particles and the Kochen-Specker contextuality theorem. Indeed, in their derivation they show that assuming relations like Eq.~\ref{1},\ref{2} (an axiom they called FIN or MIN) we can find conclusions violating the Kochen-Specker theorem (instead of Bell's theorem in the examples favored by Gisin and Blood). The details of their derivation is not useful here since it is not so different from~\cite{Gisin1,Blood}. Moreover concerning deterministic hidden variables they wrote:
\begin{quote}
\textit{It follows that there can be no correct relativistic deterministic theory of nature. In particular, no relativistic version of a hidden variable theory such as Bohm’s well-known theory can exist.}~\cite{Kochen2}
\end{quote}
The `free-will' theorem thus states that determinism is dead and that the particles are somehow `free'. This strong statement is however not necessary as we explained before and as was emphasized by Gisin and Hardy.\\
\indent On a historical ground it is thus interesting to note that Hardy's conclusions predate both Gisin-Blood's and Conway-Kochen's theorems. In~\cite{Hardy2} Hardy indeed wrote:
 \begin{quote}\textit{By using the assumption that the hidden-variable descriptions of two subsystems are disjoint when the state can be written as a product and by demanding Lorentz-invariance, we have in fact derived the locality property that the outcome of the measurement at end 1 is independent of the choice of measurement at end 2 and vice versa. Therefore, it is not surprising that we can derive a contradiction with quantum mechanics because of Bell's theorem.}~\cite{Hardy2}  
\end{quote} 
\indent Moreover, Hardy~\cite{Hardy2} obtained a more involved demonstration of the no-go theorem based on his earlier work on Hardy's nonlocality paradox without inequality~\cite{Hardy1a}. This  important deduction will  now be summarized and commented.\\ \indent In Hardy's proof we start with a two-particle  non maximally entangled state 
\begin{eqnarray}
\Psi_{H}(x_+,x_-)=\frac{1}{\sqrt{3}}(u_+(x_+)v_-(x_-)+v_+(x_+)u_-(x_-)+v_+(x_+)v_-(x_-))\label{3}
\end{eqnarray} where the particles are labeled + and - and have space-time  coordinates $x_+$, $x_-$. \begin{figure}[hbtp]
\includegraphics[width=1\textwidth]{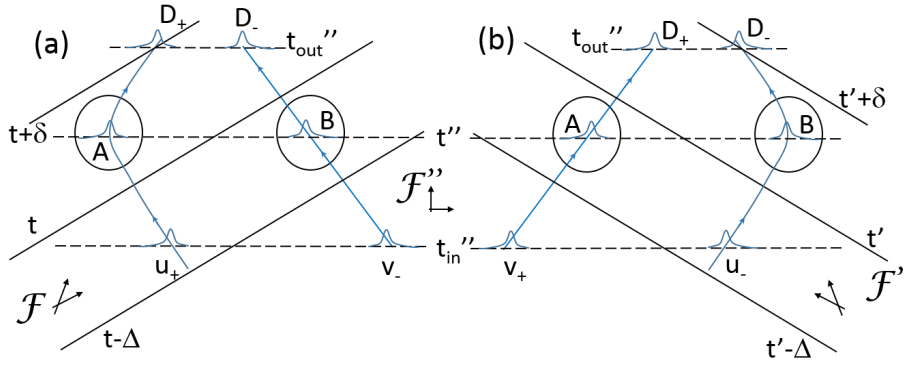}\\
\caption{ Hardy's contradiction concerning  Lorentz invariance in two Lorentz frames (a) and (b). }
\label{fig3}
\end{figure}Furthermore, as shown on Fig~\ref{fig3} the localized wave packets $u_\pm$, $v_\pm$ have no common  
spatial supports at the initial time $t_{in}''$ (defined in the Earth Lorentz frame $\mathscr{F}''$). Here we assume that the wave packets are bi-spinors associated with fermions. Eq.~\ref{3} therefore corresponds to a multi-time Dirac description of the wave function (e.g., like in the hyper-surface Bohm-Dirac (HSBD) model advocated in~\cite{Berndl,Durr}).  We also suppose that the basements of the two experiments done by Alice and Bob with particles + and - respectively  are located far away from each other so that a space-like separation can be considered in the following. At time $t''$ Alice and Bob in regions A and B use beam splitters to combine the wave packets and analyze them in the orthogonal bases  $C_\pm=U_\pm[(u_\pm+v_\pm)/\sqrt{2}]$, $D_\pm=U_{\pm}[(u_\pm-v_\pm)/\sqrt{2}]$ where $U_\pm$ are the time evolution operators acting independently for the $\pm$ particles in the multi-time formalism~\footnote{With these conventions $C_\pm(x_\pm)$ and $D_\pm(x_\pm)$ are defined after the wave packets already crossed the beam splitters.}.\\ 
\indent From Eq.~\ref{3} we can deduce~\cite{Hardy1a,Hardy2} different joint and conditional probabilities for detecting particles + and - in separated regions of space-time. First,  we have at time $t_{out}''> t''$ the joint probability 
\begin{eqnarray} 
P(D_+,D_-)=\frac{1}{12} \label{DD}
\end{eqnarray}
meaning that the probability of a joint outcome in wavepackets $D_+$ and $D_-$ is not vanishing (this formula doesn't require the  simultaneity of  $D_+$ and $D_-$ measurements).  We have also the conditional probabilities
\begin{eqnarray} 
P(u_+|D_-)=1, P(v_+|D_-)=0\label{UD}
\end{eqnarray}
imposing that a detection in $D_-$ (at any time time after $t''$) implies a detection in $u_+$ (at any times  before $t''$) if detectors are located in these regions. Symmetrically, by exchanging + and - we also  get \begin{eqnarray} 
P(u_-|D_+)=1, P(v_-|D_+)=0.\label{DU}
\end{eqnarray}       
Finally, we have at any times  before $t''$  the conditional probabilities
\begin{eqnarray} 
P(v_-|u_+)=1,  P(u_-|u_+)=0\label{VU}
\end{eqnarray}
and 
\begin{eqnarray} 
P(v_+|u_-)=1,  P(u_+|u_-)=0\label{UV}
\end{eqnarray} which means that a detection in $u_+$ implies a detection in $v_-$ while a detection in $u_-$ implies a detection in $v_+$.\\
\indent  It is key to observe that Eqs.~\ref{DD}, \ref{UD}, \ref{DU} and the couple of implications Eqs.~\ref{VU} and \ref{UV} correspond to different experimental contexts which can not be realized together and are even incompatible. A counterfactual and non contextual reasoning neglecting this fact would lead to Hardy's paradox~\cite{Hardy1a}. \\
\indent In \cite{Hardy2} the idea is to define a different contradiction for nonlocal hidden-variable theories using LIHVT and POILM. First, observe that the probabilities given in the previous equations are Lorentz invariant and represent therefore absolute facts.  Now, suppose that  for the system  of particles  + and - we consider hidden variables  $\lambda$ defined in the remote past and such that Eq.~\ref{DD} occurs.  The sub-ensemble $\Lambda_{D_+D_-}$ of such hidden variables $\lambda\in \Lambda_{D_+D_-}$ corresponds to a  subset  with probability weight  $1/12$ of the full hidden-variable space $\Lambda$ for the initial state Eq.~\ref{3} \footnote{i.e.,  $\int_{\Lambda_{D_+D_-}}d\lambda\rho(\lambda)=\frac{1}{12}$ with $\rho(\lambda)$ the normalized density of probability for $\lambda$:
 $\int_\Lambda d\lambda\rho(\lambda)=1$.}.\\ 
\indent Now, we  apply the hidden-variable model in the reference frame $\mathscr{F}$ associated with the foliation  $\mathcal{F}$  (see Fig.~3(a)) such that events in region B occur before those happening in region A. Specifically, in $\mathscr{F}$ at a time $t$ corresponding to an hyperplane $\Sigma_t\in \mathcal{F}$ the -  particle is already present in the $D_-$ wave packet while from Eq.~\ref{UD} the + particle would necessary have been detected in the $u_+$ wave packet (i.e., if a detector would have been located at the intersection  between the $u_+$ trajectory and $\Sigma_t$). Of course, in the actual experiment  there is no detector intersecting the $u_+$ beam and the particle will finish its journey in the $D_+$ (e.g., on the detector located on the hyperplane $\Sigma_{t+\delta}$).   Importantly, from POILM we know that at time $t$ the + particle doesn't have information about its future in region A. The choice to put a detector in the $u_+$ beam or instead to let the beam continues its path to the $D_+$ gate can be done at the last moment after the detection at $D_-$ occurred (i.e., admitting a possible nonlocal force acting instantaneously along  the leaves of $\mathcal{F}$).  Moreover, from Eq.~\ref{VU} applied to $\lambda\in \Lambda_{D_+D_-}$ we also know that at any time $t-\Delta$  in the past of B  we should necessarily detect the - particle in the $v_-$ beam if a detector would have been located there (which is again not the case in the actual experiment considered). Again POILM imposes a form a freedom of choice on the dynamics of the - particle provided the decision to put or not a detector in the $v_-$ beam is taken before B will occur.\\
\indent To complete the demonstration of the no-go theorem it is enough to remark that we can reproduce all the previous reasoning by using  instead of $\mathscr{F}$ a third reference frame $\mathscr{F}'$ (shown in Fig.~\ref{fig3}(b)) in which the time sequence of events is reversed  and A occurs before B. By symmetrical inferences we deduce the paths associated with the hidden-variable  space $\Lambda_{D_+D_-}$ and we realize that the result obtained  with POILM and $\mathcal{F}'$ is radically different from the previous one realized with the foliation $\mathcal{F}$.  In other words the application of POILM to a scenario like the one discussed here is strongly hyperplane or foliation dependent. This kind of trajectories are for example predicted within BM based on different folations \cite{Berndl,Durr}.  This in turn means that hidden-variable theories based on the two foliations $\mathcal{F}$ and $\mathcal{F}'$ are not equivalent and thus can not be compared using LIHVT~\cite{Hardy2}. This no-go theorem ~\cite{Hardy2} like the one by Gisin and Blood \cite{Gisin1,Blood} represents apparently a dead-end for Lorentz-invariant and deterministic hidden variable theories. Yet, POILM was conceived as a kind of Newtonian causality allowing instantaneous interactions and it is therefore not surprising to obtain a conflict with LIHVT based on Einstein's principle of relativity in space-time. To quote Hardy 
\begin{quote}
\textit{The result we have proved is analogous to the original Bell proof that hidden-variable theories are nonlocal. We have established that, with the stated conditions, they are also non-Lorentz-invariant. }~\cite{Hardy2}
\end{quote} 
As concluded by Hardy~\cite{Hardy2} and also Gisin~\cite{Gisin1} the only natural way to escape the theorem and keep hidden variables in the Minkowsky space is apparently  to provide a preferred foliation $\mathcal{F}_0$. Following Bohm~\cite{Bohm}, $\mathcal{F}_0$ is associated with a kind of Aether in which nonlocal contact could be instantaneous as in Newtonian physics. For Valentini this provides a fundamental 3+1 space-time slicing in which  the Poincar\'{e} group emerges at the statistical level~\cite{ValentiniPhd}.  In that way we restrict the application of POILM to this foliation $\mathcal{F}_0$ in order to preserve LIHVT. However, as stated by Hardy~\cite{Hardy2}, Bohm~\cite{Bohm,Ghost,Bohm2}, and Bell~\cite{Bell,Bohm}  there is no experiment that can be performed to determine what this preferred frame is (even if the absolute reference frame associated with the 2.7 K microwave cosmological background is often recalled~\cite{Bohm,Vigier2} as a good candidate in this context). Vigier and coworkers \cite{Vigier1} also proposed to use the center of mass of the full particle ensemble to define a preferred frame in BM. Similarly, Durr, Struyve et all. proposed to use the energy-tensor of a fundamental field (e.g., the Higgs field) to define a time-like and future oriented 4-vector $\langle \hat{P}^\mu \rangle_\Psi$ normal to the hyperplanes of the preferred foliation \cite{Durr2,Nomological} (a more general foliation could also be obtained by using a local time-like and future oriented fermionic current $\langle \hat{j}(x)^\mu \rangle_\Psi$ \cite{Durr2,Nomological}). Interestingly, whereas in BM  the dynamics of particles guided by the quantum state is not fully Lorentz invariant the theory is still Lorentz invariant at the statistical level meaning that the Aether is thus essentially hidden. Nevertheless, while this preferred foliation proposal mixing Newtonian and Einsteinian concepts motivates alternative theories going far beyond present-day quantum mechanics and general relativity~\cite{Vigier1,Vigier2,Bohm,Bohm2} it is difficult, or even impossible, not to see here a failure of the full deterministic hidden-variable program. In  other words, assuming an Occam razor principle most researchers would better agree that deterministic hidden variables can not be made to agree with Minkowsky space-time and should thus preferably be abandoned as an explanation of QN.
\section{Escaping  the no-go theorem with retrocausality?}\label{sec4}
\indent Introducing a preferred foliation $\mathcal{F}_0$ is however not the only strategy for amending POILM. One often neglected route is indeed to relax the constraint concerning causality and to admit backward or retro causality acting from the future to the past~\cite{Berkovitz}. This controversial solution has a very old tradition since it was already proposed in 1953 by Costa de Beauregard \cite{Costa1} to explain the EPR paradox and Bell inequalities \cite{Costa2}. Several, and sometimes even very different, quantum interpretations have in the past attempted to involve retrocausation to explain QN and EPR like correlations (e.g.,~\cite{Cramer,Gruss}) with some hidden-variable toy models \cite{Argaman,Lazarovici}. In the context of BM one can even with Sutherland develop a `Causally symmetric Bohm model' \cite{Sutherland1,Sutherland2} which is related to the two-time `teleological' interpretation proposed by Aharonov et al.\cite{Gruss} using two wave functions $\psi_i$ and $\psi_f$ associated with boundary conditions in the past and future \footnote{See also \cite{Sen} for a teleogical Bohmian model which is in fact a particular case of Sutherland model for the EPR-Bell case.}. However, it has been recently pointed out~\cite{Tumulka2,Berkovitz} that the probabilistic interpretation of such a time symmetric BM model is still in construction~and could lead to some contradictions (see also \cite{Tumulka3,Squires,Horton,Berndl} for some other BM proposals amending POILM and leading to some empirical contradictions with equivariance and Born's rule). \\      
\indent Moreover, most scientists feel reluctant for using or involving retrocausality as an explanation for QN. There are at least two good reasons for that: the first is that this is extremely counter-intuitive and against the every-day life experiences and the second is that it can leads to some contradictions like causal loops bootstrap paradoxes and so on. While the first objection is probably connected to our psychological habits grounded in the second law of thermodynamics the second is more fundamental since linked to the mathematical consistency of the theory~\cite{Berkovitz}. In this context, one of the central motivation for the preferred foliation hypothesis was given by Bell
\begin{quote}
\textit{The reason I want to go back to the idea of an Aether here is because in this EPR experiments there is the suggestion that behind the scenes something is going faster than light. Now, if all Lorentz frames are equivalent, that also means that things can go backward in time. [...] It introduces great problems, paradoxes of causality and so on. And  so it's precisely to avoid these that I want to say there is a real causal  sequence which is defined in the Aether.}~\cite{Ghost}
\end{quote} 
In other words, defining a preferred reference frame where instantaneous connections between particles is assured allow us to conceive backward causation as a mere illusion coming from us using the wrong reference frame $\mathscr{F}$ (i.e., different from $\mathscr{F}_0$). \\ 
\indent To further understand this matter about retrocausality paradoxes and causal loops we remind that space-like separated events A and B  watched from two different Lorentz reference frames can reverse time ordering and can even lead to backward signaling (see for example the perfect-crime scenario written by Bell in \cite{Bell} pp.~232-248 and invoking supraluminal particles emitted by guns). Inverting time ordering of events A and B is already what is illustrated in Fig. \ref{fig3} when we compare reference frames $\mathscr{F}$ and $\mathscr{F}'$. The idea that an effect can precede its cause is however much more demanding and is linked to faster than light signaling.\\
\begin{figure}[hbtp]
\begin{center}
\includegraphics[width=0.8\textwidth]{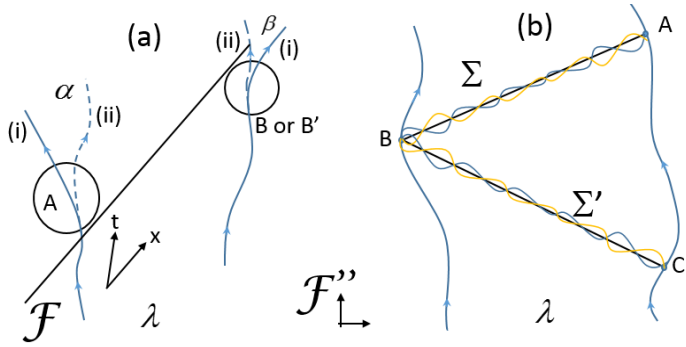}\\
\caption{ Retrocausation in quantum mechanics:  (a) Future influencing the past (b) causality loop. }
\label{fig4}
\end{center}
\end{figure}              
\indent Consider for instance the situation depicted on Fig.~\ref{fig4}(a) where (as seen from the laboratory Lorentz reference frame $\mathscr{F}''$\footnote{ From the point of view of Lorentz transformations the relation between frames $\mathscr{F}$ and $\mathscr{F}''$ in Fig.~\ref{fig4}(a) is similar to the relation existing between  frames $\mathscr{F}$ and $\mathscr{F}'$ in Figs.~\ref{fig3}(a,b).  }) two particles pass through the interaction regions A and B of space-time (with B in the future of A). Now, in the usual forward causality scenario we could imagine that  when the first particle crosses region A it emits a signal going to region B where it affects the subsequent motion of the second particle (for the moment it is irrelevant to consider whether the signal is time-like or space-like). In the backward causation scenario considered now, it is the choice made in B which affects the motion of the particle in A. For example, as shown in Fig.~\ref{fig4}(a) one could imagine that if a different experimental protocol B' was used in the region B then the motion in A and B would have switched from the trajectories labeled (i) to those labeled (ii). It is interesting to see how the observer in the lab watching this time sequence would interpret the experiment. Outcomes, $\alpha$ and $\beta$ obtained successively at A and B yield relations 
\begin{eqnarray}
\alpha=F(\Psi,\textbf{a},\textbf{b},\lambda,\lambda_A,\lambda_B),& \beta=G(\Psi,\textbf{b},\lambda,\lambda_B)\label{truc}
\end{eqnarray} 
 with $\alpha$ depending on the two settings $\textbf{a}$ and $\textbf{b}$ at A and B  whereas for the second experiment $\beta$ is independent of $\textbf{a}$ (the hidden variables $\lambda$, $\lambda_A$, $\lambda_B$ associated with the  system and the control of the settings are also included for generality). This dynamics is unusual since here the first outcome  depends on the future experiment.\\ 
\indent In classical physics such a retrocausal scenario has been already theoretically proposed  based on the time-symmetric action at a distance electrodynamics developed by Wheeler and Feynman~\cite{WF} under the name absorber theory. Importantly, in such a classical approach  the communication  channel is light-like and is used  for exchanging information through particles in region A and B with retarded and advanced solutions of Maxwell equations. It is the fine tuning  and interference between these two kinds of waves which allows the existence of backward causality and for example the possibility for an absorbing atom located in the future at B to retroactively modify the emission of an other source atom in region A.\\
\indent Solving the dynamical equations of action at a distance electrodynamics (i.e. the initial value problem) involving several particles is not however an easy task \cite{Deckert}. It causes some mathematical consistency problems as illustrated in Fig.~\ref{fig4}(b) where a particle in A interacts in a time symmetrical fashion through the link $\Sigma$ with a second particle located in B and then through a second link $\Sigma'$ retroacts back on the first particle at C, i.e., in the past of A. As emphasized by Bohm this could lead to paradoxical loops `like in the case of a person who killed his own father before he was conceived'~\cite{Bohm}. This issue is well documented in science-fiction writings exploiting time-travels  but  closed loops are not necessarily illogical or contradictory. For instance, Wheeler-Feynman action at a distance electrodynamics involving several particles makes sense if the initial data problem is not limited to a simple instantaneous Cauchy surface but involves instead knowledge about the particle motions along entire segments of trajectories \cite{Deckert}. Moreover, in the context of quantum retrocausal interpretations causal loops are not easy to remove~\cite{Berkovitz} and generate often controversies~\cite{Sutherland3,Maudlin} mixing ontological and epistemic arguments.\\
\indent Going back to BM and to the application of POILM, we find here following Bohm and Bell~\cite{Bohm,Bell} other strong arguments for a preferred foliation $\mathcal{F}_0$. Lets return to Fig.~\ref{fig4}(a) and suppose  now that A and B are space-like separated and that the system is described by an EPR wave-function $\Psi^{(-)}$ as before. According to Bell~\cite{Ghost}, the nonlocality acting in the frame $\mathscr{F}$ of Fig.~\ref{fig4}(a) (different from the previous lab frame $\mathscr{F}''$) leads to  a more usual causal reading of the experiments.  If we suppose that POILM applies to $\mathscr{F}$  then the hyperplane foliation entails a time sequence where B precedes A. The causal relations Eq.~\ref{truc} follows naturally from POILM as explained in the previous section (the function $F$ and $G$  of Eq.~\ref{truc} could thus be written as $S_{BA}$ and $F_{BA}$ in analogy Eq.~\ref{1}). In other words the nonlocal connection defined in the frame $\mathscr{F}$ is seen as retrocausation from the point of view of $\mathscr{F}''$.\\ 
\indent As we explained earlier, if LIHVT has to be preserved, this suggests the definition of a preferred frame  $\mathscr{F}=\mathscr{F}_0$ where the hidden-variable  theory will be valid and where the nonlocal connections could explain the violation of Bell inequality. However, now we see that this preferred-frame approach also yields a form of retrocausality in BM. Therefore, retrocausation in $\mathscr{F}''$  and nonlocality in $\mathscr{F}_0$ do not necessary appear as two different alternatives to POILM. For Bell this was an interesting properties of BM.\\
\indent In the same context Bohm~\cite{Bohm} considered the situation of Fig.~\ref{fig4}(b) as showing the strong plausibility of the preferred-frame picture. Indeed, if $\Sigma$ and $\Sigma'$  of Fig.~\ref{fig4}(b) are two space-like hyperplanes belonging to two different foliations $\mathcal{F}$ and $\mathcal{F}'$  we have no paradox because only one of the two foliations  can be identical to the preferred foliation $\mathcal{F}_0$ where particles interact nonlocally and where POILM applies. All this analysis, related to Gisin and Hardy no-go theorems\cite{Gisin1,Blood,Hardy2}, seems to converge in the direction of a preferred frame interpretation of hidden-variable in the Minkowsky space, and this even though the existence of such an undetectable $\mathscr{F}_0$ would violate the mere spirit of Einstein's relativity principle by introducing a new form of Aether in which instantaneous interactions are authorized.  
\section{Working with foliation-dependent and Lorentz-invariant Bohmian mechanics}                     
 \indent As we saw earlier the price to be paid for saving LIHVT is apriori very high since it entails abandoning or amending POILM by allowing retrocausation and/or the existence of a preferred frame $\mathscr{F}_0$ which mechanical description is for the moment unknown. However, if we want to respect rigorously the relativity principle  this is not the only solution for escaping the paradox and there is a much more elegant way to respect Lorentz invariance.\\ 
\indent Indeed, let suppose that in place of the preferred foliation $\mathcal{F}_0$ we introduce a statistical distribution of such foliations $\{\mathcal{F}^{(n)}_0\}$, i.e.,  an ensemble $\mathcal{F}^{(1)}_0,...,\mathcal{F}^{(n)}_0,...$ such that for each different foliation labeled by $n$ POILM applies. Now, since this is a statistical ensemble the actual system is only in one of the member $n$ of the series. Therefore,  for each actual foliation $\mathcal{F}^{(n)}_0$ every thing is like if a preferred foliation was chosen. In BM we would calculate the trajectories taking the standard guidance rules for this foliation, i.e., the trajectories become foliation of hyperplane dependent. In other words, the foliation  $\mathcal{F}^{(n)}_0$ becomes an entire part of dynamics on the same level as the wave function $\Psi$ and the hidden variables $\lambda$. For an observable outcome $\alpha$ we have thus in general
\begin{eqnarray}
\alpha=F(\Psi,\mathcal{F}^{(n)}_0, \lambda)\label{folia}\end{eqnarray}      
Importantly, as for $\lambda$ the foliation $\mathcal{F}^{(n)}_0$ is hidden meaning that a macroscopic observer (i.e., an agent) participating to an experiment has no way to access to this knowledge. However, there is an important differences between  $\lambda$ and  $\mathcal{F}^{(n)}_0$. Indeed, while the former is in BM `$|\psi|^2$' distributed along each leaf of the foliation $\mathcal{F}^{(n)}_0$ the probability distribution of foliations (i.e., in agreement with equivariance \cite{Durr}) $dP(\mathcal{F}^{(n)}_0)$ itself has no defined law. Considering an observable $\hat{A}$ we have in general the quantum average     
\begin{eqnarray}
\langle\hat{A}\rangle_\Psi=\sum_\alpha\alpha P(\alpha,\Psi)=\int\left(\int F(\Psi,\mathcal{F}_0, \lambda)\rho(\lambda,\Psi|\mathcal{F}_0)d\lambda\right) dP(\mathcal{F}_0) \label{folia2}
\end{eqnarray} where we omit the subscript $n$ and the probability element $dP(\mathcal{F}_0)$ acts on the configuration space $\mathcal{S}$ of all the foliations. Here we used Born's rule for defining $P(\alpha,\Psi)$ and the right hand part of the equation is obtained from the definition of a deterministic hidden variable~\footnote{For a deterministic dynamics we have imposed the conditional probability $P(\alpha|\lambda,\Psi,\mathcal{F}_0)=\delta_{F(\Psi,\mathcal{F}_0, \lambda),\alpha}$  (with $\delta_{i,j}$ a Kronecker symbol) which yields $\sum_\alpha \alpha P(\alpha|\lambda,\Psi,\mathcal{F}_0)=F(\Psi,\mathcal{F}_0, \lambda)$ and is taking one of the observable discrete value $\alpha$ \cite{Drezet2005}.}.\\ 
\indent Importantly, the full theory is Lorentz invariant. Whereas the choice of a foliation $\mathcal{F}_0$ clearly defines a conditioned dynamics given by Eq.~\ref{folia} the foliation distribution is a Lorentz-invariant probability measure. For foliations made of hyperplane leaves (corresponding to Lorentz frames) we can define the invariant probability measure by introducing the time-like and future oriented unit 4-vector $n_{\mathcal{F}_0}=[n_{\mathcal{F}_0}^0,\textbf{n}_{\mathcal{F}_0}]$ (i.e., with $n_{\mathcal{F}_0,\mu} n_{\mathcal{F}_0}^\mu=n_{\mathcal{F}_0}^2=1$ and $n_{\mathcal{F}_0}^0>0$) normal to the leaves $\Sigma\in\mathcal{F}_0$ (i.e., for any pair of points A and B with coordinate $x_A$ and $x_B$ belonging to a  given leaf we have $(x_A-x_B)n_{\mathcal{F}_0}=0$). The vector $n_{\mathcal{F}_0}$  completely characterizes the foliation and the probability $\delta P(\mathcal{F}_0)=\int_{\delta\mathcal{S}} dP(\mathcal{F}_0)$ on a infinitesimal set $\delta \mathcal{S}$ reads 
\begin{eqnarray}
\delta P(\mathcal{F}_0)=\int_{\delta\mathcal{S}} f(n_{\mathcal{F}_0})\delta(n_{\mathcal{F}_0}^2-1)\Theta(n_{\mathcal{F}_0}^0)d^4n_{\mathcal{F}_0}
\equiv \frac{f(\textbf{n}_{\mathcal{F}_0},\sqrt{(1+\textbf{n}_{\mathcal{F}_0}^2)})}{2\sqrt{(1+\textbf{n}_{\mathcal{F}_0}^2)}}\delta^3\textbf{n}_{\mathcal{F}_0}\nonumber\\ 
\end{eqnarray}  with  $f$ a normalized (and otherwise undefined) scalar function such that the total probability $P_{tot}=\int_{\mathcal{S}} dP(\mathcal{F}_0)=\int_{\mathbb{R}^3} \frac{f(\textbf{n}_{\mathcal{F}_0},\sqrt{(1+\textbf{n}_{\mathcal{F}_0}^2)})}{2\sqrt{(1+\textbf{n}_{\mathcal{F}_0}^2)}}d^3\textbf{n}_{\mathcal{F}_0} = 1$. While the theory doesn't give any prescription for selecting a distribution $f$  we can apriori use a principle of ignorance or indifference in a Lorentz-invariant way in order to define a microcanonical ensemble with $f=K$ where $K$ a constant. With such a choice we have   \begin{eqnarray}
P_{tot}=2\pi K\int_1 ^{+\infty} \sqrt{[(n_{\mathcal{F}_0}^0)^2-1]}dn_{\mathcal{F}_0}^0
\end{eqnarray} where  $2\pi K\sqrt{[(n_{\mathcal{F}_0}^0)^2-1]}$ acts as a probability density with respect to the variable $n_{\mathcal{F}_0}^0$. Naturally, the distribution diverges and can not be normalized (this is reminiscent of a suggestion by Dirac for defining a Lorentz-invariant Aether~\cite{Vigier1,Vigier2}) if we don't introduce a cut-off breaking Lorentz-invariance. Inversely, by selecting an infinitely narrow distribution around a particular vector $n_{\mathcal{F}_0}$ we go back to the preferred foliation approach advocated by Bohm and Bell.\\
\indent It must be emphasized that generally  speaking (i.e., if we let the probability $dP(\mathcal{F}_0)$ unspecified) the actual foliation $\mathcal{F}_0$ defines an absolute structure since it specifies a way to synchronize all the particles associated with the  guiding wave function. POILM applied to such a physical foliation is thus valid. In this dynamics the particle trajectories $x_{\mathcal{F}_0}^\mu(s)$ are thus foliation dependent. However, if instead of working in $\mathscr{F}_0$ we watch these bundle of synchronized trajectories from an arbitrary Lorentz reference frame $\mathscr{F}'$ we will not in general be able to apply POILM. It is thus very crucial to distinguish these two kinds of foliations~\footnote{Following Goldstein and Zangh\`{i} \cite{Nomological,Durr} we emphasize that any theory can be made Lorentz invariant by introducing foliations and vectors like $n_{\mathcal{F}_0}$. However, in the framework advocated here we dont want to introduce a material like absolute structure in space-time different from let say the metric tensor. Instead, foliations are parts of the integration constants for determining particle paths in BM. An analogy is provided  by the formally covariant generalization of Coulomb Gauge condition $\boldsymbol{\nabla}\cdot \mathbf{A}=0$ as $[\partial_\mu- n_\mu(n\partial)]A^\mu=0$ (with $n^2=1$) sometimes used in quantum field theory.}. It is important to emphasize the similarities and differences between the usual BM approach and the framework we propose here.    Indeed, in both approaches  the theory is fully invariant at the statistical level and the observers can not decide where is the preferred frame or the foliation $\mathcal{F}_0$. However, in the former approach  the foliation was supposed to have a material significance by adding an absolute structure to space time. In the new approach the foliation is not necessary a part of the space-time structure or of a new Aether. Better, it characterizes the particles motions by providing a synchronization over all the system (to take a Bohm analogy this is a kind of active information field) and is needed for integrating in a covariant way the quantum generalization of Hamilton-Jacobi equations.            \\
\indent The theory considered here opens several fundamental questions concerning the physical meaning  attributed  to the hidden foliations $\mathcal{F}_0$ and to the distribution $dP(\mathcal{F}_0)$. Indeed, in the preferred frame interpretation of Bohm and Bell it was implicitly stated that we should search for a mechanical basis for nonlocality  through the description of a subquantum Aether. However, since in our description we now allow for arbitrary foliation to exist the previous interpretation is not anymore acceptable. Yet, in the preferred-frame view  retrocausation was considered as a mere pathological accident coming from working in a wrong reference frame. Here this analysis is not justified anymore since LIHVT and the principle of relativity impose to consider all reference frames on an equal footing.  In that sense, the shift of paradigm is similar to the transition from the Lorentz mechanical Aether theory which motivated the research before Einstein to the modern covariant perspective in which no such a mysterious medium is needed.\\ 
\indent An analogy can be provided by comparing with the work of Goldstein and Zangh\`{i} \cite{Nomological} where the status of the wave function $\Psi$ as a guiding field is questioned and where it is suggested to interpret $\Psi$ as a nomological structure, that is, as a law-like mathematical representation without need for a material explanation.  The comparison given in \cite{Nomological} between the guiding wave function and the Hamiltonian  $H(p,q,t)$ in the phase space of classical mechanics attempts to grasp the abstract and mathematical nature of the  Bohmian description  in the configuration space as necessary (i.e., without possibility or need to return to a mechanical  interpretation in term of particles surfing on a physical wave).  A better analogy is probably between the wave function $\Psi(q,t)$ and the action $S(q,t)$ in the Hamilton-Jacobi theory which played such a fundamental role in de Broglie and Schrodinger works as well as in BM. The action is clearly an abstract representation of the possible states of motions for a specified dynamics in the configuration space. Solving the Hamilton-Jacobi equation $-\partial_t S=H(\nabla S,q,t)$ involves the finding of integral of motions. In non relativistic BM the action  plays the role of a guiding field through de Broglie formula $m \frac{dq}{dt}=\nabla S(q,t)$ where $S:=\phi$ is equivalent to the phase  $\phi$ of the wave function in the Madelung hydrodynamical representation $\Psi(q,t)=\sqrt{\rho(q,t)} e^{i\phi(q,t)}$.\\
\indent What suggests our approach however, is that the hidden foliation $\mathcal{F}_0$ is part of the integrability  problem of the Bohmian version of the Hamilton-Jacoby equation as the wave function $\Psi$ itself is. Defining a foliation $\mathcal{F}_0$ directly select a bundle of possible initial conditions satisfying the Schrodinger equation and fixing the particle dynamics at the hidden-variable level. A similar  perspective can be reached by considering the current vocabulary used in quantum foundations literature in which a separation between ontic and epistemic states are introduced \cite{PBR} (leading however to some controversies~\cite{DrezetPBR,Leifer}). The main interest of this discussion was to emphasize the role of the wave function $\Psi$ in the definition  of the hidden-variable probability space, i.e., appearing through the formula $P(\alpha,\Psi)=\int P(\alpha|\Psi,\lambda)\rho(\Psi,\lambda)d\lambda)$ for the probability of observing the eingenvalue $\alpha$ associated with the quantum observable $\hat{A}$. Here we see that $\Psi$ is present in the density of hidden variables $\rho(\Psi,\lambda)$ and in the conditional probability $P(\alpha|\Psi,\lambda)=\delta_{F(\Psi,\lambda),\alpha}$ fixing the deterministic dynamics     
(with $F(\Psi,\lambda)$ the deterministic relation for the observable).  The status of $\Psi$ and $\lambda$ is however not identical and whereas $\lambda$ is clearly a random variable defined by the choice of the initial conditions (in BM the position of the particles) oppositely $\Psi$ serves merely as a guide for the system dynamics and is (i.e. if we admit only pure quantum states) common for all particles belonging to the statistical ensemble. Now, in the new nomological framework proposed here involving both  $\Psi$ and foliations $\mathcal{F}_0$ we see (i.e., in Eq. \ref{folia2} ) that the later contributes to both the conditional dynamics as an initial condition and to the probability space as a random variable with $dP(\mathcal{F}_0)$. \\
\indent The point of view taken here is thus at minima to use a nomological interpretation of BM  and more generally of hidden  variables to develop a Lorentz-invariant primitive ontology (for the analysis of the modern notion of primitive ontology see \cite{Lam}). Going back to POILM and LIVHT we have now a way to solve the dilemma concerning the complex relation existing between the principle of relativity  and quantum mechanics. Our analysis of hyperplane or foliation dependence is limited to nonlocal and Lorentz-invariant hidden variables  but the idea is however not completely new in quantum mechanics.\\
\indent First, we remind that the notion of hypersurface dependence has a long tradition in quantum mechanics starting with Dirac work on the multitime wave function, and Tomonoga Schwinger quantum field theory of wave functional $\Psi_\Sigma$ defined on spacelike hypersurfaces (see \cite{Schweber} for a review). Second, we point out that Fleming \cite{Fleming2} analyzed the concept of hyperplane dependence in the orthodox  interpretation in order to discuss wave functions,  measurements and quantum states reduction in a covariant or relativistic way. However, the concept of wave function collapse used in the orthodox  interpretation is a non relativistic notion~\cite{Maudlin,Cohen1a} with  both epistemic and ontological contents and the proposal  of Fleming was therefore criticized by philosophers \cite{Maudlin,Maudlin2}.\\
\indent However in the present work we use the foliation dependence in the context of hidden-variable theories and the previous difficulties or vagueness with collapse do not apply. Importantly, the framework discussed here is also not completely new in the context of BM. The fundamental motivation for it came from an article by Hiley and Cohen  \cite{Cohen1a} where the idea of foliation dependence is shortly mentioned and commented~\footnote{After this work was completed I found two other works developing similar ideas \cite{Barrett,Galvan}.}. In \cite{Cohen1a} they wrote:   
\begin{quote}\textit{Another means of avoiding the requirement for a preferred frame altogether would involve abandoning the model based on a unique set of particle trajectories as a representation of the processes taking place at the beable level and replacing it with a model where, in general, the beables are represented by irreducible distributions of sets of trajectories. Each Lorentz observer would then `explicate' a single set of trajectories, defined by instantaneous nonlocal correlations in his or her particular frame. }~\cite{Cohen1a} \end{quote}   
\section{Retrocausation and nonlocality in foliation dependent Bohmian mechanics}
\label{sec5}
\indent In order to be more quantitative we will now consider some specific Bohmian models adapted to our new foliation dependent framework. We work with the primitive ontology discussed in Sec.\ref{sec1} where a system of N point-like particles piloted by a wave function  $\Psi(x_1,...,x_N):=\Psi(\{x_i\})$ is characterized at the beable level by the knowledge of the 4-coordinates $x_i(s)\in\mathbb{R}^4$ $(i=1,...N)$ . We assume that all the particle paths are labeled by a common scalar parameter $s$ playing the role of a synchronization time for the system. We postulate the existence of a covariant foliation dependent dynamics which reads 
\begin{eqnarray}
\frac{\dot{x}_i(s)}{\sqrt{\dot{x}_i(s)\dot{x}_i(s)}}=F_i(\Psi,\mathcal{F}_0,\{x_i(s)\})=\frac{J_i}{\sqrt{J_iJ_i}}(\Psi,\mathcal{F}_0,\{x_i(s)\})\label{hyper}
\end{eqnarray}           
 with $\dot{x}_i(s)=\frac{dx_i(s)}{ds}$ the parametrized  velocity. We point out that the left hand side of Eq.~\ref{hyper} is invariant through the variable change $s=f(\Psi,\mathcal{F}_0,\{x_i^{(0)}\},s')$ where $\{x_i^{(0)}\}$ is a set of initial conditions for the particles. A general solution  $x_i(s)=G_i(\Psi,\mathcal{F}_0,\{x_i^{(0)}\},s)$ can thus alternatively be written $x_i(s)=G'_i(\Psi,\mathcal{F}_0,\{x_i^{(0)}\}),s')$. The parameter $s$ is yet unspecified but we will use it to label the space-like leaves $\Sigma(s)$ of the foliation $\mathcal{F}_0$ so that all the points are on a same leaf in Eq.~\ref{hyper}. Furthermore, the foliation dependent functions $F_i$ are related to the definition of the  partial particle current $J_i$ which is defined as 
\begin{eqnarray}
J_i^{\mu_i}(\Psi,\mathcal{F}_0,\{x_i(s)\})=J^{\mu_1,...\mu_N}(\Psi,\{x_i(s)\})\Pi_{j\neq i}n_{\mathcal{F}_0,\mu_j}(x_j(s))\label{current}    
\end{eqnarray} where $J^{\mu_1,...\mu_N}(\Psi,\{x_i(s)\})$ is the foliation independent particle current obeying the local conservation rules $\partial_i J^{\mu_1,...\mu_N}(\Psi,\{x_i(s)\})=0$.\\ 
\indent To be more precise we specifically consider the HSBD model~\cite{Durr,Durr2,Tumulka2,Galvan} where the antisymmetric multi-spinor wave function for N particles  $\Psi_N(\{x_i\})\in(\mathbb{C}^4)^{\otimes N}$ is a solution of the multi-time Dirac equation~\footnote{We have $\gamma_i^{\mu_{i}}=I\otimes...\otimes\underbrace{\gamma^{\mu_{i}}}_{i^{th.} place}\otimes...\otimes I$ where $\gamma^{\mu_{i}}$ is the standard Dirac matrices. We have also  $\bar{\Psi}_N=\psi_N^\dagger\otimes_{i=1}^{i=N}\gamma_i^0$.} \begin{eqnarray}
[i\gamma_iD_i-m]\Psi_N(\{x_i\})=0\label{dirac}
\end{eqnarray} in a external electromagnetic potential $A(x)$ (i.e., with the minimal coupling $D_i=\partial_i +ieA(x_i)$,  and $e$ the electric charge).  This theory relies on the Dirac-sea or hole picture~\cite{Schweber} introduced by Bohm \cite{Bohm} where it is supposed in agreement with Dirac and the Pauli principle that in vacuum the sea of negative energy levels is filled with particles. Within this approach pair-creation is treated as a transition of a negative energy particle into a positive
energy state and real particles and antiparticles are understood as excitations and holes moving on top of the Dirac sea.\\ 
\indent The N-particle current $J$ in the HSBD model is given by 
\begin{eqnarray}
J^{\mu_1,...\mu_{N} }(\Psi,\{x_i(s)\})=\bar{\Psi}(\otimes_{i=1}^{i=N}\gamma_i^{\mu_i})\Psi\label{currentF}
\end{eqnarray} and yields a time-like and future oriented current $J_i$ defined by Eq.~\ref{current}. An interesting property of this model is that it is statistically transparent~\cite{Durr} meaning that the distribution of N path crossing the leaf $\Sigma(s)\in\mathcal{F}_0$ is given by the equivariant conserved quantity  \begin{eqnarray}
\rho_\Sigma(\{x_i(s)\})=\bar{\Psi}(\otimes_{i=1}^{i=N}\gamma_i^{\mu_i} n_{\mathcal{F}_0,\mu_i}(x_i))\Psi.\end{eqnarray} This property is however generally not true  for points not belonging to a leaf of $\mathcal{F}_0$ (a fact which has strong importance for the interpretation of quantum experiments~\cite{Durr,Lienert}).\\
\indent To illustrate the implication of such dynamical laws we go back to the examples given in Sec.~\ref{sec2} and to the no-go theorems of Gisin, Blood and Hardy~\cite{Gisin1,Blood,Hardy2}. For this  we consider a two-particle entangled system like the one described by the EPR singlet $\Psi^{(-)}$ or the Hardy quantum state Eq.~\ref{3} and write $x_i(s):=[t_i(s),X_i(s)]$ ($i=1,2$) the  space-time positions associated with the two particles for a 1D motion. The two stations  A and B of Sec.~\ref{sec2} (where measurements on particle 1 and  particle 2 respectively  occur) are located far apart from each other and in the Earth Lorentz reference frame $\mathscr{F}''$ they are separated by the typical distance $\delta X''=L$. Now, given a  different Lorentz frame $\mathscr{F}$ and  an hyperplane foliation $\mathcal{F}_0:=\mathcal{F}$ (like the one shown in Fig.~\ref{fig1} or \ref{fig3}(a)) we can  use the local time $s:=t$ in $\mathscr{F}$ for labeling the 2-path.  From Eq.~\ref{hyper} applied in $\mathscr{F}$ yields the characteristic equations 
\begin{eqnarray}
\frac{dX_1}{dt}(t)=\frac{F^1_1}{F^0_1}(\Psi,\mathcal{F},X_1(t),t,X_2(t),t)\nonumber\\
\frac{dX_2}{dt}(t)=\frac{F^1_2}{F^0_2}(\Psi,\mathcal{F},X_1(t),t,X_2(t),t)
\label{hyper2} 
\end{eqnarray} where we used the condition $t_1=t_2$ in this reference frame. This dynamics reminiscent from nonrelativistic BM is strongly non local and particles 1 and 2 are instantaneously in touch through the quantum potential~\cite{Bohm}. Moreover, from the Lorentz transformation relating the two Lorentz frames $\mathscr{F}''$ and  $\mathscr{F}$ we can rewrite the dynamics in the Earth laboratory~\footnote{ We have $t=(t''-v x'')/\sqrt{1-v^2}$ where $v<1$ is the relative velocity between the frames and thus $t_1=t_2=s$ implies $t''_1=t''_2-v(X''_2-X''_1)\simeq t''_2-vL$. }:
\begin{eqnarray}
\frac{dX''_1}{dt''}(t''-\varepsilon)=\frac{F''^1_1}{F''^0_1}(\Psi,\mathcal{F},X''_1(t''-\varepsilon),t''-\varepsilon,X''_2(t''),t'')\nonumber\\
\frac{dX''_2}{dt''}(t'')=\frac{F''^1_2}{F''^0_2}(\Psi,\mathcal{F},X''_1(t''-\varepsilon),t''-\varepsilon,X''_2(t''),t'')
\label{hyper3} 
\end{eqnarray}  where $\varepsilon=vL$, $v$ is the constant velocity difference between the two reference frames and where the position of the particle 2 and 1 appearing in the dynamics are defined at two different local times $t''_2:=t''$ and $t''_1=t''-vL$. The issue here was to refer to the foliation dependent particle motions $x''_i(s_i)$ in frame $\mathscr{F}''$ by introducing several running parameters $s_i:=t''_i$ associated with local times instead of working with the common parametrization $s$. However, since these times are not independent we have a synchronization between particles involving surpraluminal signaling and even retrocausation (as it is clearly visible from the arguments of functions in Eq. \ref{hyper3} where motion of particle 1 at time $t''-\varepsilon$ is affected by the motion of particle 2 at later time $t''$).\\
\indent Another, but yet completely pertinent, Bohmian dynamics could be obtained by using a different foliation $\mathcal{F}_0:=\mathcal{F}'$ associated with the reference frame $\mathscr{F}''$ obtained by Lorentz transformation in the opposite direction  (i.e., deduced after the change $v\rightarrow -v$  in Eq. \ref{hyper3}). This would immediately leads to 
\begin{eqnarray}
\frac{dX'_1}{dt'}(t')=\frac{F'^1_1}{F'^0_1}(\Psi,\mathcal{F}',X'_1(t'),t',X'_2(t'),t')\nonumber\\
\frac{dX'_2}{dt'}(t')=\frac{F'^1_2}{F'^0_2}(\Psi,\mathcal{F}',X'_1(t'),t',X'_2(t'),t')
\label{hyper2b} 
\end{eqnarray}
where time $t'$ defines a new synchronization between particles acting nonlocally and instantaneously in the frame $\mathscr{F}'$. In the Earth laboratory frame this alternatives dynamics reads 
\begin{eqnarray}
\frac{dX''_1}{dt''}(t'')=\frac{F''^1_1}{F''^0_1}(\Psi,\mathcal{F}',X''_1(t''),t'',X''_2(t''-\varepsilon),t''-\varepsilon)\nonumber\\
\frac{dX''_2}{dt''}(t''-\varepsilon)=\frac{F''^1_2}{F''^0_2}(\Psi,\mathcal{F}',X''_1(t''),t'',X''_2(t''-\varepsilon),t''-\varepsilon)
\label{hyper3b} 
\end{eqnarray} which is very similar to Eq. \ref{hyper3} but with the role of particle 2 and 1 exchanged concerning retrocausation.\\
\indent  In a general experiment involving entangled particles the nonlocality will induces different evolution for different synchronization foliations like $\mathcal{F}$ and $\mathcal{F}'$ and subsequently different alternative N-path congruences (and with Eq.~\ref{folia} different quantum observable histories) will be obtained. Yet, at the statistical level all the foliations $\mathcal{F}_0$  are available (even though hidden to observers) and are weighted by the probability $dP(\mathcal{F}_0)$ (see Eq.~\ref{folia2}).  In this perspective, the no-go theorems \cite{Gisin1,Blood,Hardy2,Kochen1,Kochen2} discussed in 
Sec.~\ref{sec2} result from not seeing that the contradicting histories like those obtained in Figs.~\ref{fig1},\ref{fig2} and \ref{fig3} are associated with different realizations of the hidden variable dynamics corresponding to different possible choices for foliations $\mathcal{F}_0$, i.e., for boundary/initial conditions.\\         
\begin{figure}[hbtp]
\begin{center}
\includegraphics[width=0.8\textwidth]{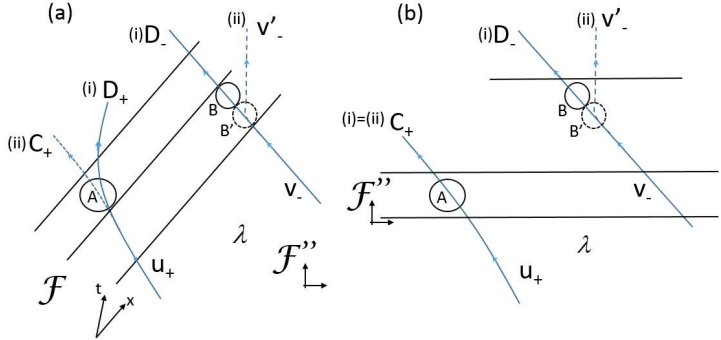}\\
\caption{ Bohmian quantum nonlocality and retrocausality with different foliations (a) and (b).  }
\label{fig5}
\end{center}
\end{figure}  
\indent The implications of the new framework on retrocausality are particularly startling. Going back to the situation depicted in Fig.~\ref{fig4}(a) where regions A and B are spacelike separated we see that an action at B or B' can retroact on the outcome $\alpha$ in region A in agreement with Eq.~\ref{truc}. However, this effect is not directly exploitable to create a `future influencing the past machine' (otherwise this could be used to modify the past). There is here a form of no-signaling theorem which prohibits us to exploit this in a quantum experiment. \\  
\indent To illustrate this point  we consider a version of Hardy's paradox shown in Fig.~\ref{fig5}(a) where particles prepared at the hidden variable level in the $u_+$, $v_-$ gates are escaping  in the  $D_+$, $D_-$ gate after interacting with beam splitters in region A and B (this presupposes the foliation $\mathcal{F}_0=\mathcal{F}$  where  B occurs before A for defining BM and corresponds to the path labeled (i) in Fig.~\ref{fig5}(a).  From Sec.\ref{sec2} (i.e. Eq.~\ref{DD}) and \cite{Hardy1a,Hardy2} we know that this event occurs with the  joint probability $P(D_+,D_-)=1/12$. We remind that this is true for $\lambda\in \Lambda_{D_+D_-}\subset\Lambda_{u_+v_-}$  where $\Lambda_{u_+v_-}$ is the hidden-variable subspace for particles in the $u_+,v_-$ gates. Moreover, if instead of using the beam splitter B we put a detector or a mirror in B' (at a time $t_{B'}<t_B$ in frame $\mathscr{F}$) then the - particle will end up in the prolongation $v'_-$ of the $v_-$ gate while the + particle (prepared in the $u_+$ state) will necessarily end up in the $C_+$ gate (as shown in Fig.~\ref{fig5}(a) with the trajectories labeled (ii)). This will occurs for all particle pairs prepared in $\lambda\in \Lambda_{D_+D_-}$. \\ 
\indent We emphasize that more generally for any two-particle entangled states characterized by a deterministic hidden-variable dynamics the probability to get the outcomes values $\alpha$, $\beta$, with the measurement settings $\textbf{a}$, and $\textbf{b}$ in region A and B conditioned on hidden variables $\lambda$, the quantum state $\Psi$, and the foliation $\mathcal{F}_0$ is:
 \begin{eqnarray}
P(\alpha,\textbf{a},\beta,\textbf{b}|\lambda,\Psi,\mathcal{F}_0)=P(\alpha,\textbf{a}|\beta,\textbf{b},\lambda,\Psi,\mathcal{F}_0)P(\beta,\textbf{b}|\lambda,\Psi,\mathcal{F}_0)\nonumber\\
=\delta_{S_{BA}(\Psi,\mathcal{F}_0, \textbf{b}, \textbf{a}, \lambda),\alpha}\delta_{F_{BA}(\Psi,\mathcal{F}_0, \textbf{b}, \lambda),\beta}
\label{proba2} 
\end{eqnarray}
 This probability can only takes value 0 or 1 and depends on the observable functions as defined in Eq.~\ref{truc}.  The causal structure associated with POILM is clearly visible since $F_{BA}$ doesnt depend on the later choice to be done on the $\textbf{b}$ settings while $S_{BA}$ depends nonlocally on both $\textbf{a}$, $\textbf{b}$ settings.\\ 
\indent   
In the present case illustrated in Fig.~\ref{fig5}(a) we have for $\lambda\in \Lambda_{D_+D_-}$ and from Eq. \ref{proba2} the following list of conditionnal probabilities~\footnote{For $\lambda\in \Lambda_{D_+D_-}$ we have $P_{B}(D_+|\lambda)=P(D_+|D_-,\lambda)P(D_-|\lambda)+P(D_+|C_-,\lambda)P(C_-|\lambda)=1+0=1$ and $P_{B'}(C_+|\lambda)=P(C_+|v'_-,\lambda)P(v'_-|\lambda)+P(C_+|u'_-,\lambda)P(u'_-|\lambda)=1+0=1$. The other probabilities are similarly obtained.} \begin{eqnarray}
P_{B}(D_+|\lambda)=1, &  P_{B}(C_+|\lambda)=0 \nonumber\\
P_{B'}(D_+|\lambda)=0, & P_{B'}(C_+|\lambda)=1
\label{proba1} 
\end{eqnarray}  
where $P_{B}(D_+|\lambda)$ (respectively $P_{B}(C_+|\lambda)$) means the conditional probability for detecting the + particle in exit $D_+$ (respectively $C_+$) in region A knowing that a second beam splitter is located in B and that $\lambda\in \Lambda_{D_+D_-}$. Similarly  $P_{B'}(D_+|\lambda)$, $P_{B'}(D_+|\lambda)$ imply that  detectors are located in region B'. Nothing of causally surprising here and everything results from POILM  and QN applied to $\mathcal{F}$.\\ 
\indent However, if we  now analyze the problem from the lab reference frame $\mathscr{F}''$  where A occurs before B  then we have clearly retrocausation since the choice to put the detector in B' or to work with the beam splitter in B is delayed after the + particle ever crossed the region A!  Therefore we end up with a scenario where the future can influence the past. Observe however that in these hidden variable models $\rho(\lambda)$ is not modified in the remote past because it is here the dynamics which is modified through backward and faster than light signaling (this contrasts with other retrocausality approaches~\cite{Argaman}). Here retrocausation is driven  by faster than light signaling `propagating' along the leaves $\Sigma\in\mathcal{F}$ \\
\indent Yet, this retrocausation will only occur if the foliation $\mathcal{F}_0=\mathcal{F}$ used for defining the dynamics given by Eqs.~\ref{hyper2},\ref{hyper3} is such that A occurs before B and B'.  For a different choice of integration constants associated with a different foliation (e.g., $\mathcal{F}''$ in Fig.~\ref{fig5}(b) there is no such a backward causality. In  the case of Fig.~\ref{fig5}(b) the particle pair is still prepared in the $u_+$, $v_-$ gate. For both scenario (i) and (ii) interaction (as seen from $\mathscr{F}''$) occurs at A first and the + particle will necessarily end up in gate $C_+$. The choice at B or B' is done later and according to POILM can not affect the probability of outcomes $C_+$ or $D_+$ at A. In the scenario (i) the - particle can go to $C_-$ or $D_-$ (on the figure the - particle goes to $D_-$).  Moreover, if scenario (ii) occurs, i.e. if a detectors are located at B', we will still have the + particle ending its journey in $C_+$ gate while  the - particle starting in $v_-$ will necessarily end up in the $v'_-$ gate. in other words, for  $\lambda\in \Lambda_{u_+v_-}$ we have $P_{B}(C_+|\lambda)=P_{B'}(C_+|\lambda)=1$ and $P_{B}(D_+|\lambda)=P_{B'}(D_+|\lambda)=0$ \footnote{A detailed analysis shows that $P_{B}(C_+|\lambda)=P_{B'}(C_+|\lambda)$ and $P_{B}(D_+|\lambda)=P_{B'}(D_+|\lambda)$ for any $\lambda\in \Lambda$. }. This alternative BM doesn't show backward in time reaction.\\
\indent The previous example exploits POILM and Hardy's paradox to get retrocausation for some specific values of hidden variables such that  
$\lambda\in \Lambda_{D_+D_-}$ for the foliation $\mathcal{F}_0=\mathcal{F}$. However BM shows that this result is much more general and robust and that in fact it could affects all particle pairs of the statistical ensemble $\lambda\in \Lambda$ for the foliation $\mathcal{F}_0=\mathcal{F}$.  We could for example consider the case proposed by Rice \cite{Rice,Bricmont} in which two particles are prepared in the EPR state 
\begin{eqnarray}
\Psi_{R}(x_+,x_-)=\frac{1}{\sqrt{2}}(u_+(x_+)u_-(x_-)+v_+(x_+)v_-(x_-))\label{EPRrice}
\end{eqnarray} where as in  previous examples the beams associated with $u_\pm(x_\pm)$ and  $v_\pm(x_\pm)$ have disjoint space-time supports before the space-like separated regions A and B where local measurements occur. Now instead of a beam splitter in A and B we simply let the beams $u_+(x_+)$ and  $v_+(x_+)$ (respectively $u_-(x_-)$ and  $v_-(x_-)$) crossed each other in A (respectively B) to end up in exit $C_+$ and  $D_+(x_+)$ (respectively $C_-$ and  $D_-$). Nothing of special in the usual interpretation of quantum mechanics: we record a perfect correlation between the detections at $D_+$ and $D_-$  or between  the detections at $C_+$ and $C_-$, i.e., $P(D_+,D_-)=P(C_+,C_-)=1/2$. However, from BM different things happen because the dynamics given by Eqs. \ref{hyper}, \ref{hyper2} is first order in time and forbid two paths in the configuration space to cross each other \cite{Rice,Bricmont}. If we consider a foliation $\mathcal{F}_0=\mathcal{F}$ such that A and B are simultaneous the only solution as pointed by Bricmont is that the paths followed by the particles bounce off each other. The + particle starting in the $u_+$ ($v_+$) beam will thus end up in the $D_+$ ($C_+$) gate. The same is true for the - particle. Moreover, if a mirror is located in region B or if detectors are included in B' the - particle beams $u_-(x_-)$ and  $v_-(x_-)$ will now end up in the $u'_-(x_-)$ and  $v'_-(x_-)$ exit gates. This corresponds to a `which-path' for the  + particle and Bohmian paths are now allowed to cross: the particle starting in the $u_+$ ($v_+$) beam will thus end up in the $C_+$ ($D_+$) gate. Importantly, since A and B/B' are space like separated we can analyze the problem from a different referent frame $\mathscr{F}''$ in which A occurs before B.  Again, we have retrocausation since by using BM with the foliation $\mathcal{F}_0=\mathcal{F}$ the choice made at B to use or not a mirror is seen from $\mathscr{F}''$ as following the experiment at A while the causal dynamics imposed by BM imposes that B causes A.  Also, this will clearly occurs for any particle-pair of the ensemble.   Yet, like for the previous examples a different choice of foliation $\mathcal{F}_0=\mathcal{F}''$ would prohibit retrocausality.  In the present case this will happen as soon as we consider a foliation where  A and B are not simultaneous and retrocausation will thus be avoided for all particles of the ensemble. \\
\indent Therefore, we see that retrocausation is not easy to isolate.  Since both $\lambda$ (i.e. the particle coordinates) and the foliations  $\mathcal{F}_0$ are hidden  the macroscopic observer has in general no possibility to exploit retrocausality for modifying or shaping the past in a observable way (even though quantum particles do retroact at the fundamental level). This is a form of no-signaling theorem which could be evaded if a preferred frame was selected \textit{and} if quantum equilibrium was broken. For example retrocausality could be observed and used to change the past if we would be able to prepare (i.e., with a kind of Bohmian version of Maxwell's demon \cite{Valentini}) the system  in the state $\lambda\in\Lambda_{D_+D_-}$ with the foliation $\mathcal{F}_0=\mathcal{F}$ as indicated in Fig.~\ref{fig5}(a). The passage of the configuration (i) to (ii) decided at B/B' located later than A could thus modify the earlier dynamics at A (note that the remote past before A would not be modified) and this could be observed at the statistical level. In other words we could build up a kind of Bohmian time machine...\\
\indent However, in a state of quantum equilibrium  an observer at A can only measure a local probability $P(C_+)=5/6$ or $P(D_+)=1/6$ irrespectively of what is occurring at B/B' and of the foliation $\mathcal{F}_0$ used to define the particle dynamics: a time machine is statistically prohibited and our usual causality protected from backward causation.\\
\indent It is quite remarkable that such foliation-dependent deterministic hidden-variable models afford QN together with retrocausality. In turn, it also shines some new light on old discussions concerning delayed-choice quantum-erazer and entanglement-swapping experiments which are going back to debates between Heisenberg, von Weizsacker, Einstein and Hermann (for a complete review see~\cite{Ma}). In these delayed-choice experiments entanglements between a meter and a particle leads ultimately to a EPR-like scenario like the one illustrated on Fig.~\ref{fig4}(a) where stations A and B are space-like separated and with $t''_A<t''_B$ in the  Lorentz frame considered.   All our previous analyzes  apply to such scenarios and therefore retroaction indeed occurs for some foliations  $\mathcal{F}_0$ in agreement with intuitions concerning delayed-choice quantum-erazer. Still, all these experiments are constrained by our no-signaling theorem prohibiting to exploit such retrocausation at the statistical level to modify the past (i.e., in the limit of the quantum equilibrium conditions).\\        
\indent A final remark on the issue of retrocausation concerns causal loops. The kind of contradiction obtained in Fig.~\ref{fig4}(b) is in the new framework clearly forbidden because it would require to use two foliations (with leaves $\Sigma\in\mathcal{F}$ and $\Sigma'\in\mathcal{F}'$) at once in order to get the loop between point A, B and C. Since foliations $\mathcal{F}$ and $\mathcal{F}'$ are not actualized at the same time for the same particle there is no contradiction. The foliation dependent HSBD model advocated here is thus immune to such paradoxes.    
\section{Possible generalizations, the limit of the nomological interpretation and a conclusion}
\label{sec6}
\indent Results discussed in Sec. \ref{sec5} were based on the particle HSBD ontology developed in~\cite{Durr,Durr2,Tumulka2}. Yet, this is certainly not the only way to apply the new framework for foliation dependent hidden-variable theories.\\
\indent A possible extension concerns BM applied to bosonic quantum field. Following the Schr\"{o}dinger-functional approach advocated by Bohm \cite{Bohm,Cohen1a} (for reviews see \cite{Holland,Struyve}) a bosonic quantum field is described by the wave-functional $\Psi([ \phi(x)]_\Sigma)$ where  $[\phi(x)]_\Sigma$ denotes the set of bosonic field values $\phi(x)$ at each point $x$ of the leaf $\Sigma\in \mathcal{F}$. The beables in this version of BM are not point-like particles but the field values $\phi(x)$ on the leaves of the foliation.\\ 
\indent The most known example is the scalar real field obeying to the Klein-Gordon equation and which is characterized for a foliation $\mathcal{F}_0$ by the guidance equation~\cite{HortonB,Oldstein,Durr2} 
\begin{eqnarray}
\frac{D}{Dt}\phi(x)|_\Sigma:=n_{\mathcal{F}_0}(x)\partial \phi(x)=\frac{\delta S([ \phi(x)]_\Sigma) }{\delta_\Sigma \phi(x)} 
\label{klein}
\end{eqnarray}
 where  $S([ \phi(x)]_\Sigma) $  is the phase of $\Psi([ \phi(x)]_\Sigma)$ and $\frac{\delta  }{\delta_\Sigma \phi(x)}$  is a (foliation dependent) functional derivative operator  \footnote{For a functional $G([ \phi(x)]_\Sigma)$ and a function $f(x)$ we have $\int_{\Sigma} d^3\sigma(x)f(x)\frac{\delta G([ \phi(x)]_\Sigma) }{\delta_\Sigma \phi(x)}=\lim_{\varepsilon\rightarrow 0} \frac{ G([ \phi(x)+\varepsilon f(x)]_\Sigma)- G([ \phi(x)]_\Sigma)}{\varepsilon} $ with $d^3\sigma(x)$ an elementary invariant hypersurface \cite{Durr2}.}. We sketch a derivation of this covariant Bohmian quantum-field ontology with its main properties in the appendix.  Importantly, this model is statistically transparent and respect equivariance of Born's rule for the probability density $|\Psi([ \phi(x)]_\Sigma)|^2$ on the leaves of $\mathcal{F}_0$. Moreover, this ontology for quantum field is highly nonlocal (i.e., much more than the usual Bohmian ontology for particles \cite{Bohm}). In particular,  states with fixed number of bosons are not localized in space-time and therefore the ontology is counter-intuitive. Yet, it leads to observable
 consequences agreeing with standard quantum mechanics when coupled with localized fermionic detectors \cite{Dewdney} (different minimalistic versions of the theory exist in the literature \cite{Struyve2,Gold}). Retrospectively, such Bohmian models of bosons interacting with localized fermions reintroduce a form of (deterministic \footnote{We note that within our foliation dependent framework one could easily develop  a generalization of the GRW stochastic spontaneous collapse\cite{GRW} approach in a way different from Tumulka's. For this purpose one could consider a stochastic choice of the foliation $\mathcal{F}_0$ which would actualize one foliation over a distribution $dP(\mathcal{F}_0)$. The rest of GRW \cite{GRW} written in a given foliation $\mathcal{F}_0$ would be kept unchanged.} ) wave-function collapse different from the GRW-flash proposal~\cite{Tumulka,GRW}. A foliation dependent framework applied to such an ontology would imply that for each foliation $\mathcal{F}_0$ the dynamical law Eq. \ref{klein} involves a different alternative evolution of the quantum field $\phi(x):=\phi_{\mathcal{F}_0}(x)$. Nonetheless, this is is fully Lorentz-invariant if we consider foliations $\mathcal{F}_0$ as integrating constants of the pilot-wave dynamics.  Due to the specificity of this model QN and retrocausation will appear in the infinite-dimensional configuration space  of fields (regularized in some ways to avoid infrared and ultraviolet divergences~\cite{Struyve}). The approach is very promising in the context of quantum gravity  where the most natural beables are the components $g^{\mu\nu}(x)$ of the metrical tensor~\cite{Holland,ValentiniPhd}. However, for quantum particles like mesons or photons an ontology based on fields creates a strong asymmetry with fermions (e.g., leptons or quarks) described by HSBD particle models. The relations with the non relativistic Bohm particle model existing for both fermions and bosons is also difficult to clarify in this perspective.  It is possible to introduce a foliation dependent framework for relativistic bosonic particles (e.g., obeying to the Klein-Gordon equation) but this will be discussed in a subsequent article.  \\ 
\indent The different examples discussed here  illustrate the general idea  of our foliation-dependent framework for deterministic hidden-variable. We propose the introduction of foliation dependent beables  $X_{\mathcal{F}_0}(s)$ where the different variables are synchronized through the foliation $\mathcal{F}_0$  with leaves $\Sigma(s)\in\mathcal{F}_0$ parametrized by $s$. The recipe is relatively safe and can be summarized like that: (i) take any relativistic Bohmian ontology discussed in the literature. All these models involve a preferred foliation. (ii) then take a statistical ensemble or mixture  of systems prepared with different foliations but identical state initial state $|\Psi_{\Sigma_{in}}\rangle$ defined on a space-like hyper-surface. The new theory is empirically equivalent to quantum mechanics at the statistical level. Yet, the idea respects the spirit of serious Lorentz-invariance, i.e., Einstein relativity principle. Here, contrarily to previous proposals \cite{Bohm,Bell,ValentiniPhd,Vigier1,Durr2,Gold}  the choice of a foliation $\mathcal{F}_0$ is  not fixing a new Aether or a 3+1 slicing of space-time providing an absolute standard of simultaneity. Rather, it defines integrating constants for determining the quantum evolution of the beables. In the approach advocated by Bohm, Hiley~\cite{Bohm} and Valentini~\cite{ValentiniPhd} the Poincar\'{e} group is conceived as an emerging symmetry valid at the statistical level in the regime of quantum equilibrium.  However, here we restore the fundamental symmetry: there is no Aether and every Lorentz frames are equivalent in agreement to LIHVT.  The foliation $\mathcal{F}_0$ represents the actual synchronization between some internal clocks associated with particles (or elementary volumes for a field ontology). With this actual synchronization POILM holds and QN defines a causal connection acting from past to future (in the frame $\mathscr{{F}_0}$). Crucially, during an experiment the observer repeat many times a procedure including preparation of a wave function $\Psi_{\Sigma_{in}}$ and detection of localized events. However, he or she has no control on the way the particles or fields are synchronized. Furthermore, the foliation choice can fluctuate randomly between different runs of the same experiment implying that at the statistical level quantum mechanics is preserved (i.e., in the quantum equilibrium regime where nonlocality and retrocausality are hidden). \\ 
\indent In this work we fully supported a nomological approach where  wave functions $\Psi_{\Sigma(s)}$ on the foliations $\mathcal{F}_0$ are seen as an abstract dynamics without need for a deeper description (a bit like in the older classical Hamilton-Jacobi formalism which motivated the work by de Broglie). This is a form of minimalistic primitive ontology which can be used for computing paths and trajectories in the Minkowsky space. The theory takes seriously a block-universe perspective and foliations are not as physical as in the approach advocated by Bohm Hiley or Bell with the preferred frame (our approach is thus more in harmony with the ideas of general relativity putting all reference frames on a same footing). Clearly, for our limited purpose of calculating paths we don't have to know if the wave-function with its foliation $\mathcal{F}_0$ (acting in the configuration space) is a form of material wave interacting with particles or fields.\\ 
\indent However, I do not think that a such a nomological view~\cite{Nomological} constitutes an happy end for the story. Indeed, QN is completely  described by our hidden-variable approach but is it really explained? Bohmian mechanics (e.g. the foliation dependent HSBD model) is a remarkable example since  the guidance condition is postulated as a principle  without further explanation (de Broglie himself never accepted this conclusion and devoted most of his energy to explain and derive this law). The introduction of foliations adds an other level of abstraction which, like QN, cries for an explanation. Furthermore, since  QN  now comes out with retrocausality  this stresses even more the peculiar nature of quantum mechanics in the Minkowsky space.  Admittedly,  backward in time causation, i.e., future influencing the past opens indeed fundamental questions concerning the notion of free-will and on the notion of super-determinism (i.e., conspiratorial common cause in the past) abhorred by Bell \footnote{As bell wrote \textit{apparently separate parts of the world would be deeply and conspiratorially entangled, and our apparent free will would be entangled with them.} \cite{Bell}, p. 154. }. The notion of super-determinism  or causal conspiracy has indeed an intricate relation with retrocausation. Take for example the well-known Wheeler-Feynman absorber theory~\cite{WF} in which electromagnetic fields $F^{\mu\nu}$ are essentially time-symmetric and can be expanded has half retarded and half advanced source fields:  \begin{eqnarray}F^{\mu\nu}=\frac{F_{ret.}^{\mu\nu}+F_{adv.}^{\mu\nu}}{2}.\label{WF1}\end{eqnarray} Now, from usual Green's theorem we can express any such a field equivalently as a sum of initial and retarded components $F_{in}^{\mu\nu}+F_{ret.}^{\mu\nu}$ or as a sum of final and advanced components  $F_{out}^{\mu\nu}+F_{adv.}^{\mu\nu}$(with $F_{in/out}^{\mu\nu}$ free fields). Here we have \begin{eqnarray}F_{in}^{\mu\nu}=-F_{out}^{\mu\nu}=\frac{F_{adv.}^{\mu\nu}-F_{ret.}^{\mu\nu}}{2}.\label{WF2}\end{eqnarray} This freedom in the expansion of fields has a consequence since it allows an observer watching time from past to future to interpret Wheeler-Feynman's theory  from a more usual retarded perspective. In order to do that the observer has however to include a conspiratorial field $F_{in}^{\mu\nu}$ given by Eq.~\ref{WF2} to the usual retarded field $F_{ret.}^{\mu\nu}$. In this view something of very peculiar happening in the remote past determines the initial boundary conditions to produce the nice conspiracy or miracle needed in the Wheeler-Feynman approach. In that sense retrocausation is a special case of super-determinism and we are apparently free, if we wish, to reinterpret the `future influencing the past' links by using our usual causality, i.e., after adding some dose of super-determinism or conspiracy in the initial boundary conditions. Such common causes would look miraculous and antithermodynamical (i.e., unprobable) for an observer watching the time flowing from past to future. Yet, in turn it would explain and debunk retrocausation.\\
\indent  In foliation-dependent Bohmian mechanics we have also retrocausation  and therefore the natural question  which arises is whether a super-deterministic and fatalistic interpretation should not be possible as well to understand QN in Bohmian mechanics? With the present nomological interpretation (and specially within the particle primitive ontology advocated in the HSBD model where particles have no internal structure in the Minkowsky space) this is not possible since the fundamental object is the wave function which is used to define the highly nonlocal guidance equations. With the nomological interpretation there is no (local) substructure equivalent to the electromagnetic field in the Wheeler-Feynman approach and which could be used to give us a mechanical explanation of QN and retrocausation. Therefore, it lets the door open for new original propositions going beyond Bohmian mechanics in order to conciliate  QN, retrocausation and super-determinism. To end-up with a provocative but yet optimistic sentence I can only quote Bell who once wrote
\begin{quote}
\textit{I am quite convinced of that: quantum mechanics is only a temporary expedient.}~\cite{Ghost}
\end{quote} 
\begin{acknowledgements}                              
I thank C\'{e}dric Poulain, Cyril Branciard, Vincent Lam, and Jean Bricmont for helpful discussions and comments. 
\end{acknowledgements}
\section{Appendix: A foliation dependent Bohmian ontology for Bosonic quantun fields}
\label{secAP}
\indent We follow \cite{Long} and use the Schr\"{o}dinger wave-functional picture\footnote{Our description mathematically extends an earlier result by Valentini~\cite{ValentiniPhd} (obtained with $N=1$ in Eq.~\ref{ADM}) but with a completely different physical interpretation since we don't here advocate a preferred 3+1 foliation of space-time.}.  For this purpose we consider in the Minkoswky flat space-time  (as seen from a Lorentz frame with metric $\eta_{\mu\nu}$) the classical action $S=\int d^4x \mathcal{L}(\phi(x),\partial\phi(x),x)$ for a real scalar field $\phi(x)$. This action is equivalently analyzed using the curvilinear coordinate system $x':=[s,\xi^i]$ ($i=1,2,3$) with the transformation $x^\mu=x^\mu(s,\xi^i)$ defined such that $s$ labels the leaves $\Sigma(s)\in\mathcal{F}_0$.\\
\indent Following the ADM formalism~\cite{ADM} we define a `lapse' function $N(x')$ and three tangential projections $p^\mu_i$ such that 
\begin{eqnarray} 
dx^\mu=Nn^\mu ds+p^\mu_id\xi^i
\label{ADM}
\end{eqnarray} and $p^\mu_in_\mu=0$ with $n^\mu$ the vector normal to the leaf  $\Sigma(s)$ at point  of coordinate $x^\mu$ (we use our freedom in the choice of coordinates to cancel the `shift' function $N^i=0$ \cite{Long,ADM}).
 With ADM notations we thus have $\frac{\partial x^\mu}{\partial s}=N n^\mu$, $\frac{\partial x^\mu}{\partial \xi^i}=p^\mu_i$, $\frac{\partial s}{\partial x^\mu}=\frac{n_\mu}{N}$ characterizing the coordinate transformation.
  Writing   $g'_{\mu\nu}$ the metric in the $x'$ coordinate system we deduce $g'_{00}=N^2$, $g'_{0i}=g'_{i0}=0$, $g'_{ij}=h_{ij}=\eta_{\mu\nu}p^\mu_i p^\nu_i$ and $\sqrt{-g'}=N\sqrt{-h}$ with $g'(x')$, and $h'(x')$ the determinant of $g'_{\mu\nu}$ and $h_{ij}$ respectively.\\
\indent The action $S$ for the field $\phi(x)=\phi'(s,\xi)$ reads now with $\mathcal{L}(\phi(x),\partial\phi(x),x)=\mathcal{L'}(\phi'(s,\xi),\partial_s\phi'(s,\xi),\nabla_i\phi'(s,\xi),s,\xi)$:
\begin{eqnarray}
S=\int dsd^3\xi N\sqrt{-h}\mathcal{L'}(\phi'(s,\xi),\partial_s\phi'(s,\xi),\nabla_i\phi'(s,\xi),s,\xi)\label{Action}
\end{eqnarray} with $\nabla_i$ a short hand notation for $\frac{\partial}{\partial \xi^i}$, $\nabla_i\phi'=p^\mu_i\partial_\mu \phi$ and $\partial_s\phi'=N n^\mu\partial_\mu \phi$. Of course, Euler-Lagrange's equation $\partial_s(\sqrt{-g'}\frac{\partial \mathcal{L'}}{\partial \partial_s\phi'})+\nabla_i(\sqrt{-g'} \frac{\partial \mathcal{L'}}{\partial \nabla_i\phi'})=\sqrt{-g'}\frac{\partial \mathcal{L'}}{\partial\phi'}$ deduced from Eq. \ref{Action} is rigorously equivalent to $\partial_\mu\frac{\partial\mathcal{L}}{\partial\partial_\mu\phi}=\frac{\partial \mathcal{L}}{\partial\phi}$ obtained in the $x$ coordinate system in agreement with general relativistic covariance.\\
\indent Writing $S=\int dsL_{\Sigma(s)}$ we use a Legendre transformation to define the Hamiltonian  as $H_{\Sigma(s)}a=-L_{\Sigma(s)} +\int d^3\xi \Pi'\partial_s\phi'$ with $\Pi'=\frac{\delta L_\Sigma}{\delta \phi'}=\sqrt{-g'}\frac{\partial \mathcal{L'}}{\partial \partial_s\phi'}=\sqrt{-h}\frac{\partial \mathcal{L}}{\partial \partial_\mu\phi}n_\mu$ the canonical momentum conjugate to $\phi$~\footnote{ Here we use  the usual definition of the functional derivative: for a functional $G([ \phi'(s,\xi)])$ and a function $f(x)=f'(s,\xi)$ we have $\int_{\Sigma} d^3\xi f'(s,\xi)\frac{\delta G([ \phi'(s,\xi)]) }{\delta \phi'(s,\xi)}=\lim_{\varepsilon\rightarrow 0} \frac{ G([ \phi'(s,\xi)+\varepsilon f'(s,\xi)])- G([ \phi'(s,\xi)])}{\varepsilon} $. This definition is different from the covariant one used in footnote 14  and involving the invariant elementary hypersurface $d^3\sigma= d^3\xi\sqrt{-h}$.  We have $\frac{\delta G([ \phi(x)]_\Sigma) }{\delta_\Sigma \phi(x)}=\frac{1}{\sqrt{-h}}\frac{\delta G([ \phi'(s,\xi)]) }{\delta \phi'(s,\xi)}$}. This entails
\begin{eqnarray}
H_{\Sigma(s)}=\int d^3\xi N\sqrt{-h}(\frac{\partial \mathcal{L'}}{\partial \partial_s\phi'}\partial_s\phi'-\mathcal{L'})=\int_\Sigma d^3\sigma(x) N\mathcal{H}_\Sigma(x)
\end{eqnarray} with $d^3\sigma= d^3\xi\sqrt{-h}$  and $\mathcal{H}_\Sigma(x)=\mathcal{H'}_\Sigma(x')$ a foliation dependent scalar energy density such that $\mathcal{H}_\Sigma=T^{\mu\nu}n_\mu n_\nu$ with $T^{\mu\nu}=\frac{\partial \mathcal{L}}{\partial \partial_\mu\phi}\partial^\nu \phi-\eta^{\mu\nu}\mathcal{L}$ the full energy-momentum tensor\footnote{ $T^{\mu\nu}(x)$ satisfies the conservation law $\partial_\mu T^{\mu\nu}=-\partial^\nu \mathcal{L}|_{\phi,\partial\phi}$ where the explicit derivative holds for the explicit $x$ dependence in $\mathcal{L}$ in presence of external fields.}. In the Hamilton formalism we have explicitly $\mathcal{H'}_\Sigma=\mathcal{H'}_\Sigma(\phi',\pi'(s,\xi),\nabla_i\phi',x')$ where $\pi'_\Sigma=\frac{\Pi'}{\sqrt{-h}}=\frac{\partial \mathcal{L}}{\partial \partial_\mu\phi}n_\mu$ is introduced for further convenience.\\
\indent In order to quantize this theory  we introduce the equal-time commutation relations $[\hat{\phi'}(s,\xi),\hat{\phi'}(s,\xi')]=0=[\hat{\Pi'}(s,\xi),\hat{\Pi'}(s,\xi')]=0$ and
\begin{eqnarray}
[\hat{\phi'}(s,\xi),\hat{\Pi'}(s,\xi')]=i\delta^3(\xi-\xi')
\end{eqnarray}
written in the generalized Heisenberg picture adapted to the foliation where $s$ plays the role of a time parameter\footnote{Moreover, the Covariance of the dynamics is better appreciated when using the canonical momentum $\hat{\pi_\Sigma'}=\frac{\hat{\Pi'}}{\sqrt{-h}}$ leading to the commutation relation  $[\hat{\phi'}(x'),\hat{\pi_\Sigma'}(y')]=i\delta_\Sigma^3(x,y)$ for $x,y\in \Sigma $. $\delta^3_\Sigma(x,y)$ is a Dirac distribution such that for $x,y\in \Sigma $ we have $\delta^3_\Sigma(x,y)=\delta^3_\Sigma(y,x)=\frac{\delta^3(\xi_x-\xi_y)}{\sqrt{-h(x')}}$ and therefore $\int_\Sigma d^3\sigma f(x)\delta^3_\Sigma(x,y)=f(y)$ if $x\in\Sigma$.}. The relation with the Schr\"{o}dinger picture involves an unitary transformation such that for any local operator in the Heisenberg picture $\hat{A}^{(H)}(x):=\hat{A}([\hat{\phi'}(s,\xi),\hat{\Pi'}(s,\xi)],s)$ it exists a Schr\"{o}dinger representation $\hat{A}^{(S)}(x):= \hat{A}([\hat{\phi'}(s_{in},\xi),\hat{\Pi'}(s_{in},\xi)],s)$ 
\begin{eqnarray}\hat{A}^{(H)}(x)=\hat{U}_{\Sigma(s),\Sigma_{in}(s_{in})}^{-1}\hat{A}^{(S)}(x)\hat{U}_{\Sigma(s),\Sigma_{in}(s_{in})}\label{Heisewave}\end{eqnarray}
 where $s_{in}$ labels an initial leaf $\Sigma(s_{in})\in\mathcal{F}_0$ \footnote{In Eq.~\ref{Heisewave} if $\hat{A}$ depends explicitly on $s$ this label is not modified between the two pictures. This is is the case for the Hamiltonian density $\mathcal{H}_\Sigma(x)=T^{\mu\nu}(x)n_\mu(x) n_\nu(x)$.}. The wave functional at time $s$ (Schr\"{o}dinger picture) is related to the one at time $s_{in}$ (Heisenberg picture) by $|\Psi_{\Sigma(s)}\rangle=\hat{U}_{\Sigma(s),\Sigma_{in}(s_{in})}|\Psi_{\Sigma_{in}(s_{in})}\rangle$ with the Schr\"{o}dinger equation:
\begin{eqnarray}
i\frac{d}{ds}|\Psi_{\Sigma(s)}\rangle=\int_\Sigma  d^3\sigma N\mathcal{\hat{H}'}_\Sigma(\hat{\phi'}(s_{in},\xi),\nabla\hat{\phi'}(s_{in},\xi),\frac{\hat{\Pi'}}{\sqrt{-h}}(s_{in},\xi),s,\xi)|\Psi_{\Sigma(s)}\rangle\label{Schrowave}\nonumber\\
\end{eqnarray} We emphasize that the Hamiltonian is here written in the Schr\"odinger picture.\\
\indent 
To work with the Schr\"{o}dinger functional representation we introduce the amplitude 
\begin{eqnarray}
\Psi([\phi'(\xi)],s)=\langle[\phi'(\xi)];s|\Psi_{\Sigma_{in}(s_{in})}\rangle=\langle[\phi'(\xi)];s_{in}|\Psi_{\Sigma(s)}\rangle
\end{eqnarray}with the eigenvectors condition $\hat{\phi'}(s,\xi)|[\phi'(\xi)];s\rangle=\phi'(\xi)|[\phi'(\xi)];s\rangle$ and the evolution 
$|[\phi'(\xi)];s\rangle=\hat{U}_{\Sigma(s),\Sigma_{in}(s_{in})}^{-1}|[\phi'(\xi)];s_{in}\rangle$. 
Furthermore, we have the representation  
\begin{eqnarray}
\langle[\phi'(\xi)];s_{in}|\hat{\Pi'}(\xi,s_{in})|\Psi(s)\rangle=-i\frac{\delta \Psi([\phi'(\xi)],s)}{\delta \phi'(\xi)}
\end{eqnarray} which yields
\begin{eqnarray}
i\frac{\partial}{\partial s}\Psi([\phi'(\xi)],s)=\int_\Sigma  d^3\sigma N\mathcal{\hat{H'}}_\Sigma(\phi'(\xi),\nabla\phi'(\xi),\frac{-i}{\sqrt{-h}}\frac{\delta  }{\delta \phi'(\xi)},s,\xi)\Psi([\phi'(\xi)],s)\label{SchrowaveB}
\nonumber\\
\end{eqnarray}
\indent Moreover, BM entails the introduction of field beables defined on $\Sigma(s)\in\mathcal{F}_0$. The foliation dependent formalism advocated here imposes thus the beables $\phi'(\xi):=\phi'_{\mathcal{F}_0}(x')=\phi_{\mathcal{F}_0}(x)$ for points $x\in \Sigma(s)$ and allows us to write
\begin{eqnarray}
\Psi([\phi'(\xi)],s):=\Psi([\phi_{\mathcal{F}_0}(x)]_{\Sigma(s)})
\end{eqnarray}
The fundamental Schr\"odinger wave-functional equation reads now 
\begin{eqnarray}
i\frac{\partial}{\partial s}\Psi([\phi_{\mathcal{F}_0}]_{\Sigma(s)})=\int_\Sigma  d^3\sigma N\mathcal{\hat{H'}}_\Sigma(\phi_{\mathcal{F}_0},\nabla\phi_{\mathcal{F}_0},-i\frac{\delta  }{\delta_\Sigma\phi_{\mathcal{F}_0}},x')\Psi([\phi_{\mathcal{F}_0}]_{\Sigma(s)}) \label{42}
\nonumber\\
\end{eqnarray}
(see footnotes 14 for notations).
 As an example we consider the field described classically  by the Lagrangian $\mathcal{L}=\frac{1}{2}\partial_\mu\phi\partial^\mu \phi- V(\phi)$ leading to the quantum Hamiltonian density (in the Schr\"{o}dinger representation) $\mathcal{\hat{H'}}_\Sigma=\frac{\hat{\pi'}_\Sigma^2}{2}-\frac{h^{ij}}{2}\nabla_i\hat{\phi'}\nabla_i\hat{\phi'}+V(\hat{\phi'})$. Eq.\ref{42} entails  
\begin{eqnarray}
i\frac{\partial}{\partial s}\Psi([\phi_{\mathcal{F}_0}]_{\Sigma})=\int_\Sigma  d^3\sigma N[\frac{-\delta^2  }{2\delta_\Sigma\phi^2_{\mathcal{F}_0}}-\frac{h^{ij}}{2}\nabla_i\phi_{\mathcal{F}_0}\nabla_i\phi_{\mathcal{F}_0}+V(\phi_{\mathcal{F}_0})]\Psi([\phi_{\mathcal{F}_0}]_{\Sigma}) \label{43}
\nonumber\\
\end{eqnarray}
The Madelung polar expansion $\Psi([\phi_{\mathcal{F}_0}]_{\Sigma(s)})=R([\phi_{\mathcal{F}_0}]_{\Sigma(s)})e^{iS([\phi_{\mathcal{F}_0}]_{\Sigma(s)})}$ leads to the Bohmian Hamilton-Jacobi equation 
\begin{eqnarray}
-\frac{\partial}{\partial s}S=\int_\Sigma  d^3\sigma N[\frac{1}{2}\left(\frac{\delta S }{\delta_\Sigma\phi_{\mathcal{F}_0}}\right)^2-\frac{h^{ij}}{2}\nabla_i\phi_{\mathcal{F}_0}\nabla_i\phi_{\mathcal{F}_0}+V(\phi_{\mathcal{F}_0})]+Q_\Sigma\nonumber\\
  \label{44}
\end{eqnarray} involving the quantum potential $ Q_\Sigma=-\int_\Sigma  d^3\sigma N\frac{1}{2R}\frac{\delta^2 R }{\delta_\Sigma\phi^2_{\mathcal{F}_0}}$, and the probability conservation
 \begin{eqnarray}
-\frac{\partial}{\partial s}R^2=\int_\Sigma  d^3\sigma N \frac{\delta  }{\delta_\Sigma\phi_{\mathcal{F}_0}}\left(R^2\frac{\delta S }{\delta_\Sigma\phi_{\mathcal{F}_0}}\right)\label{45}
\end{eqnarray} from which we derive the probability conservation $\int \mathcal{D}\phi_{\mathcal{F}_0}R^2(s)=1$ ($\mathcal{D}\phi_{\mathcal{F}_0}$ is a functional volume~\cite{Hatfield} defined in the configuration space at time $s$).\\
\indent Most importantly, BM is driven by the guidance equation 
\begin{eqnarray}
\pi'_\Sigma=\frac{\delta S }{\delta_\Sigma\phi_{\mathcal{F}_0}(x)}=Im\left(\frac{1}{\Psi}\frac{\delta \Psi }{\delta_\Sigma\phi_{\mathcal{F}_0}(x)}\right)=n_{\mathcal{F}_0}^\mu(x)\partial_\mu\phi_{\mathcal{F}_0}(x)=\frac{\partial_s\phi_{\mathcal{F}_0}(x)}{N}
\end{eqnarray}
 which is equivalent to Eq.~\ref{klein} discussed in \cite{HortonB,Oldstein,Durr2}.
While BM is clearly a first-order dynamics we can yet deduce the Newton-like second-order differential equation by applying the functional derivative on both sides of Eq.~\ref{44}. It yields:  
\begin{eqnarray}
\partial_\mu\partial^\mu\phi_{\mathcal{F}_0}(x)=\frac{1}{\sqrt{-g'}}[\partial_s(\frac{\sqrt{-h}}{N}\partial_s\phi'_{\mathcal{F}_0}(x'))+\nabla_i(h^{ij}\nabla_j\phi'_{\mathcal{F}_0}(x'))]\nonumber\\
=-\frac{dV(\phi)}{d\phi}|_{\phi=\phi_{\mathcal{F}_0}(x)}-\frac{1}{N}\frac{\delta Q_\Sigma }{\delta_\Sigma\phi_{\mathcal{F}_0}(x)}\label{47}
\end{eqnarray}
which differs from the classical equation $\partial_\mu\partial^\mu\phi_{\mathcal{F}_0}(x)=-\frac{dV(\phi)}{d\phi}|_{\phi=\phi_{\mathcal{F}_0}(x)}$ by the introduction of a nonlocal  and foliation dependent quantum force responsible for the `super-implicate order' advocated by Bohm and Hiley \cite{Bohm}.\\
\indent We emphasize that while we actually picked up a specific foliation $\mathcal{F}_0$ for representing the Schr\"odinger wave-functional problem, the full structure is still entirely relativistically covariant. To see this, we introduce the general transformation $|\Psi_{\Sigma'}\rangle=\hat{U}_{\Sigma',\Sigma}|\Psi_{\Sigma}\rangle$ where $\Sigma, \Sigma'$ do not necessarily belong to $\mathcal{F}_0$. Let $x_\Sigma^\mu$ be any point of $\Sigma(s)\in\mathcal{F}_0$. We then define an infinitesimal variation of the surface $\Sigma(s)\rightarrow\Sigma'$  by the transformation $x_\Sigma^\mu\rightarrow x_\Sigma^\mu+\epsilon n^\mu(x_\Sigma)$ where $\epsilon(\xi)$ is the infinitesimal and local amount of displacement normal to $\Sigma$. The unitary infinitesimal transformation relating $\Psi([\phi]_{\Sigma'})$ and $\Psi([\phi]_{\Sigma(s)})$ leads to 
\begin{eqnarray}
\Psi([\phi]_{\Sigma'})-\Psi([\phi]_{\Sigma(s)})\simeq -i\int_\Sigma  d^3\sigma \epsilon\mathcal{\hat{H'}}_\Sigma(\phi,\nabla\phi,-i\frac{\delta  }{\delta_\Sigma\phi},x')\Psi([\phi]_{\Sigma(s)}) 
\nonumber\\
=\int_\Sigma  d^3\sigma \epsilon\frac{\delta}{\delta\Sigma(x)}\Psi([\phi]_{\Sigma(s)})\label{48}
\end{eqnarray} where we introduced in the second line the definition of Schwinger's functional derivative $\frac{\delta}{\delta\Sigma(x)}\Psi([\phi]_{\Sigma(s)})$~\cite{Doplicher,Schweber}. From this we deduce
a multi-time  Schwinger-Tomonaga equation~\cite{Tomonaga} adapted to the Schr\"{odinger}-Heisenberg picture~\cite{Matthews} 
\begin{eqnarray}
i\frac{\delta}{\delta\Sigma(x)}\Psi([\phi]_{\Sigma})=\mathcal{\hat{H'}}_\Sigma(\phi,\nabla\phi,-i\frac{\delta  }{\delta_\Sigma\phi},x')\Psi([\phi]_{\Sigma}) 
\label{49}
\end{eqnarray}
which connects with the Bohmian description given in \cite{HortonB,Oldstein,Durr2}. We emphasize that $[\mathcal{\hat{H'}}_\Sigma(x_1),\mathcal{\hat{H'}}_\Sigma(x_2)]=0$ $\forall x_1,x_2\in \Sigma$ as it should be in this formalism~\cite{Tomonaga,Schweber,Matthews}.


\begin{thebibliography}{}
\bibitem{Bell}
Bell, J.S.: Speakable and unspeakable in quantum mechanics, second edition. Cambridge University Press, Cambridge (2004).
\bibitem{deBroglie} Bacciagaluppi, G. and Valentini, A.: Quantum theory at the crossroads: Reconsidering the 1927 Solvay Conference. Cambridge University Press, Cambridge (2009).  
\bibitem{Bohm} Bohm., D and Hiley, B.J.: The undivided Universe. Routledge, London, (1993).
\bibitem{Bell2} Bell, J.S.: unpublished interview by R. Weeber (1990).  
\bibitem{ValentiniPhd}
Valentini, A.: On the pilot-wave theory of classical, quantum and subquantum physics. International School for advanced studies, Trieste (1992).
\bibitem{Ghost} Davies, P.C. and Brown, J.R.: The Ghost in the atoms, Chaps. 3, 8, 9. Cambridge University Press, Cambrigde (1986).
\bibitem{Vigier1} Kyprianidis, A., Vigier, J.P.: Quantum action-at-a-distance: the mystery of Einstein-Podolsky-Rosen correlations. In: Selleri, F. (ed.) Quantum mechanics versus local realism. Springer, New York (1988).  
\bibitem{Vigier2} Combourieu, M.-C., and Vigier, J.~P.:  Absolute space-time and realism in Lorentz invariant interpretations of quantum mechanics.  Phys. Lett. A \textbf{175}, 269 (1993).   
\bibitem{Hardy1a} Hardy, L.:  Quantum mechanics, local realistic theories and Lorentz-invariant realistic theories. Phys. Rev. Lett. \textbf{68}, 2981 (1992).
\bibitem{discussion} Berndl, K., and Goldstein, S.: Comment on ``Quantum mechanics, local realistic theories and Lorentz-invariant realistic theories''.  Phys. Rev. Lett. \textbf{72}, 780 (1994).
\bibitem{Cohen1a}
Cohen, O., and Hiley, B.J.: Reexamining the assumption that elements of reality can be Lorentz invariant. Phys. Rev. A \textbf{52}, 76 (1995).
\bibitem{Hardy2}Hardy, L.  and Squires, E.J.: On the violation of Lorentz-invariance in deterministic hidden-variable interpretations of quantum mechanics. Phys. Lett. A \textbf{168}, 169 (1992).
\bibitem{Kochen1}
Conway, J. and Kochen, S.: The free will theorem. Found. Phys. \textbf{56}, 1441 (2006).
\bibitem{Kochen2}
Conway, J.  and Kochen, S.: The strong free will theorem. Notices of the American Mathematical Society \textbf{66}, 226 (2009).
\bibitem{Gisin1}
Gisin, N.:  Impossiblity of covariant deterministic nonlocal hidden-variable extension of quantum theory. Phys. Rev. A \textbf{83}, 020102(R) (2011).
\bibitem{Gisin2}
Gisin, N.: The free will theorem, stochastic quantum dynamics and true becoming in relativistic quantum physics (2010).  arxiv:1002.1392.
\bibitem{Blood}
Blood, C.: Derivation of Bell's locality condition from the relativity of simultaneity (2010). arxiv:1005.1656.
\bibitem{discussion2a}
Tumulka, R.: Comment on ``the free will theorem''. Found. Phys. \textbf{37}, 186 (2007).
\bibitem{discussion2b} 
Conway, J. and Kochen, S.: Reply to comments of Bassi, Ghirardi, and Tumulka on the free will theorem. Found. Phys. \textbf{37}, 1643 (2007).
\bibitem{discussion2c}
Goldstein, S., Tausk, D.V., Tumulka R. and Zanghi, N.: What does the free will theorem actually prove? (2009). arxiv:0905.4641v1. 
\bibitem{Tumulka}
Tumulka, R.: A relativistic version of the Ghirardi-Rimini-Weber model. J. Stat. Phys. \textbf{125}, 821 (2006).
\bibitem{GRW}
Ghirardi, G.C., Rimini, A., Weber, T.: Unified dynamics for microscopic and macroscopic systems. Phys. Rev. D  \textbf{34}, 470 (1986).
\bibitem{Bricmont}
Bricmont, J.:  Making sense of quantum mechanics, Chap.~5, pp. 162-169, Springer International Publishing, Switzerland (2016).
\bibitem{Rice} Rice, D.A.: A geometric approach to nonlocality in the Bohm model of quantum mechanics. Am. J. Phys. \textbf{65}, 144 (1997).
\bibitem{Berndl} M\"{u}nch-Berndl, K., D\"{u}rr, D., Goldstein, S., Zangh\`{i}, N.: Nonlocality, Lorentz invariance, and Bohmian quantum theory. Phys.~Rev.~A \textbf{53}, 2062 (1996).
\bibitem{Durr}  D\"{u}rr, D., Goldstein, S., M\"{u}nch-Berndl, K.,Zangh\`{i}, N.: Hypersurface Bohm-Dirac models. Phys.~Rev.~A \textbf{60}, 2729 (1999).
\bibitem{Bohm2}
Bohm, D. and Hiley, B.J.: On the relativistic invariance of a quantum  theory based on beables. Found. Phys. \textbf{21}, 243 (1991).
\bibitem{Durr2}  
 D\"{u}rr, D., Goldstein, S., Norsen, T., Struyve, W., Zangh\`{i}, N.: Can Bohmian mechanics be made relativistic. Proc.~R.~Soc. A \textbf{470}, 20130699 (2014).
\bibitem{Nomological}
Goldstein, S., and Zangh\`{i}, N.: in The wave function: essays in the metaphysics of quantum mechanics, Albert, D. and  Ney, A. eds., Oxford University Press (2012).
\bibitem{Berkovitz}
Berkovitz, J.: On Predictions in Retro-causal Interpretations of Quantum Mechanics. Stud. Hist. Philos. Mod. Phys. \textbf{39}, 709 (2008).
\bibitem{Costa1}
Costa de Beauregard, O.: Une r\'{e}ponse \`{a }l’argument dirig\'{e} par Einstein, Podolsky et Rosen contre l’interpr\'{e}tation bohrienne des ph\'{e}nom\`{e}nes quantiques.  C. R. Acad. Sci. Paris \textbf{236}, 1632 (1953).
\bibitem{Costa2}
Costa de Beauregard, O.: Lorentz and CPT invariances and the Einstein-Podolsky-Rosen correlations. Phys. Rev. Lett \textbf{50}, 867 (1983).
\bibitem{Cramer}
Cramer, J.G.: The transactional interpretation of quantum mechanics. Rev. Mod. Phys. \textbf{58}, 647 (1986).
\bibitem{Gruss}
Aharonov, Y., Gruss, E.Y.: Two-time interpretation of quantum mechanics. e-print arXiv:quant-ph/0507269
\bibitem{Argaman}
Argaman, N.: Bell’s theorem and the causal arrow of time. Am. J. Phys. \textbf{78}, 1007 (2010).
\bibitem{Lazarovici}
Lazarovici, D.: A relativistic retrocausal model violating Bell's inequality. Proc.~R.~Soc. A \textbf{471}, 20140454 (2014).
\bibitem{Sutherland1}
Sutherland, R.I.: Causally symmetric Bohm model. Stud. Hist. Philos. Mod. Phys. \textbf{39}, 782 (2008).
\bibitem{Sutherland2}
Sutherland, R.I.: Lagrangian description for particle interpretations of quantum mechanics: entangled many-particle case. Found. Phys \textbf{47}, 174 (2017).
\bibitem{Sen}
Sen, I.: A local $\psi$-epistemic retrocausal hidden-variable model of Bell correlations with wavefunctions in physical space. Found. Phys. \textbf{XX}, (2018).
\bibitem{Tumulka2}
Tumulka, R.: On Bohmian mechanics, particle creation, and relativistic space-time: happy 100th Birthday, David Bohm!   Entropy \textbf{20}, 462 (2018). 
\bibitem{Tumulka3}
Goldstein, S., Tumulka, R.: Opposite arrows of time can reconcile relativity and nonlocality. Class. Quant. Grav. \textbf{20}, 557 (2003). 
\bibitem{Squires}
Squires, E.J.: A local hidden-variable theory that FAPP, agrees with quantum theory.  Phys. Lett. A 178, 22 (1993).
\bibitem{Horton}
Horton, G., Dewdney, C.: Nonlocal, Lorentz-invariant, hidden variable interpretation of quantum mechanics. J. Phys. Math. Gen. \textbf{34}, 9871 (2001). 
\bibitem{WF}
Wheeler, J.A. and Feynman, R.P.: Interaction with the absorber as the mechanism of radiation.  Rev. Mod. Phys \textbf{17}, 157 (1945). 
\bibitem{Deckert}
Deckert, D.-A.: Electrodynamics absorber theory-a mathematical study,  Der Andere Verlag (2010).
\bibitem{Sutherland3}
Sutherland, R.I.: A corollary to Bell's theorem. Il Nuovo Cimento \textbf{88B}, 114 (1985).
\bibitem{Maudlin}
Maudlin, T.: Quantum non-locality and relativity. Blackwell, Oxford (1994).
\bibitem{Drezet2005}
Drezet, A.: Comment on ``A simple experiment to test Bell’s inequality", J.-M. Vigoureux. Opt. Commun. \textbf{250}, 370 (2005).  
\bibitem{PBR}
Pusey; M.F., Barrett, J., Rudolph, T.: Nature Physics  \textbf{8}, 475 (2012).
\bibitem{DrezetPBR}
Drezet, A.: On the reality of the quantum state. Int. J. Quantum Found. \textbf{1}, 25 (2015). 
\bibitem{Leifer}
Leifer, M.S.: Is the Quantum State Real? An Extended Review of $\psi$-ontology Theorems. Quanta \textbf{3}, 68 (2014). 
\bibitem{Lam}
Lam, V.: Primitive ontology and quantum field theory. Euro. Jnl. Phil. Sci.  \textbf{5}, 387 (2015).
\bibitem{Schweber}
Schweber, S.S.: QED and the men who made it, Princeton University Press, Princeton (1994).
\bibitem{Fleming2} 
Fleming, G., Bennett, H.: Hyperplane dependence in relativistic quantum mechanics. Found. Phys. \textbf{19}, 231 (1989).
\bibitem{Maudlin2}
Maudlin, T.: Space-time in the quantum world. In: Cushing, J.T., Fine, A., Goldstein, S. (eds.) Bohmian mechanics and quantum theory an appraisal, pp. 285-307. Kluwer academic publishers (1996). 
\bibitem{Barrett}
Barrett, J.A.: Relativistic Quantum Mechanics through Frame--Dependent Constructions. Philosophy of Science \textbf{72}, 802 (2005). 
\bibitem{Galvan}
Galvan, B.: Relativistic Bohmian mechanics without a preferred foliation. J. Stat. Phys. \textbf{161}, 1268 (2015).  
\bibitem{Colin}
Colin, S., Struyve, W.: A Dirac sea pilot-wave model for quantum field theory. J. Phys. A \textbf{40}, 7309 (2007). 
\bibitem{Lienert}
Lienert, M., Tumulka, R.: Born's Rule for Arbitrary Cauchy Surfaces. e-print arXiv:1706.07074v2 
\bibitem{Valentini} Valentini, A.: Signal-locality, uncertainty, and the subquantum H-theorem II. Phys. Lett. A \textbf{158}, 1 (1991).
\bibitem{Ma}
Ma, X.S., Koffer, J., Zeilinger, A.: Delayed-choice gedanken experiments and their realizations. Rev. Mod. Phys. \textbf{88}, 015005 (2016).
\bibitem{Holland}
Holland, P.H.: The de Broglie-Bohm theory of motion and quantum field theory. Phys. Rep. \textbf{224}, 95  (1993). 
\bibitem{Struyve} Struyve, W.: Pilot-wave theory and quantum fields. Rep. prog. phys. \textbf{73}, 106001 (2010). 
\bibitem{HortonB} Horton, G., Dewdney C.: A relativistically covariant version of Bohm's quantum field theory for the scalar field.  J. Phys. A: Math. Gen. \textbf{37}, 11935 (2004).
\bibitem{Oldstein}
D\"{u}rr, D., Goldstein, S., Zangh\`{i}, N.: On a realistic theory for quantum physics. In: Albeverio,S., Casati,G., Cattaneo, U., Merlini, D.v (eds.)  Stochastic processes, physics and geometry, pp. 374-391. World Scientific, Singapore  (1990).
\bibitem{Dewdney}   Dewdney, C., Horton, G.,  Lam, M.M., Malik, Z., Schmidt, M.: Wave-particle dualism and the interpretation of quantum mechanics, Found. Phys. \textbf{22}, 1217 (1992). 
\bibitem{Struyve2} Struyve, W.: Pilot-wave approaches to quantum field theory. J. Phys.: Conf. Ser. \textbf{306}, 012047 (2011). 
\bibitem{Gold} Goldstein, S., Taylor, J., Tumulka, R., Zangh\`{i}, N.: Are all particles real?  Stud. Hist. Philos. Mod. Phys. \textbf{36}, 103 (2005).
\bibitem{Long}
Long, D.V., Shore, G.M.: The Schr\"{o}dinger wave functional and vacuum states in curved spacetime. Nucl. Phys. B \textbf{530}, 247 (1998).
\bibitem{ADM}
Arnowitt, R., Deser, S., Misner, C.: Dynamical Structure and Definition of Energy in General Relativity. Phys. Rev. \textbf{116}, 1322 (1959). 
\bibitem{Hatfield}
Hatfield, B.: Quantum field theory of point particles and strings. Addison-Wesley, Redwood City (1992). 
\bibitem{Doplicher}
Doplicher, L.: Generalized Tomonaga-Schwinger equation from the Hadamard formula. Phys. rev. D. \textbf{70}, 064037 (2004). 
\bibitem{Tomonaga}
Tomonaga, S. On a relativistically invariant formulation of the quantum theory of wave fields. Prog. Phys.\textbf{ 1}, 27 (1946).
\bibitem{Matthews}
Matthews, P.T.: The generalized Schr\"odinger equation in the interaction representation Phys. Rev. \textbf{75}, 1270 (1949) 
\end{thebibliography}
\end{document}